\documentclass[useAMS,usenatbib]{mn2e}
\usepackage{graphicx,natbib,color,multirow,amsmath,epstopdf}
\def\lesssim{\mathrel{\hbox{\rlap{\hbox{\lower3pt\hbox{$\sim$}}}\hbox{\raise2pt\hbox{$<$}}}}}
\def\gtrsim{\mathrel{\hbox{\rlap{\hbox{\lower3pt\hbox{$\sim$}}}\hbox{\raise2pt\hbox{$>$}}}}}
\begin{document}

\title[Thermal and Kinetic SZE of Merging Clusters]{The Observable Thermal and Kinetic Sunyaev-Zel'dovich Effect in Merging Galaxy Clusters}

\author[J.~J.~Ruan, T.~R.~Quinn, and A.~Babul]{John~J.~Ruan$^{1}$\thanks{E-mail:
jruan@astro.washington.edu}, Thomas~R.~Quinn$^{1}$ and Arif~Babul$^{2}$\\
$^{1}$Department of Astronomy, University of Washington, Box 351580, Seattle, WA, 98195, USA\\
$^{2}$Department of Physics and Astronomy, University of Victoria, Victoria, BC, V8P 1A1, Canada}

\date{Accepted 2000 December 01. Received 2000 December 01; in original form 2000 October 01}

\pagerange{\pageref{firstpage}--\pageref{lastpage}} \pubyear{2000}

\maketitle
\label{firstpage}

\begin{abstract}
The advent of high-resolution imaging of galaxy clusters using the Sunyaev-Zel'dovich Effect (SZE)
provides a unique probe of the astrophysics of the intracluster medium (ICM) out to 
high redshifts. To investigate the effects of cluster mergers on resolved SZE images, we 
present a high-resolution cosmological simulation of a 1.5$\times$10$^{15}$ M$_{\odot}$ 
adiabatic cluster using the TreeSPH code ChaNGa. This massive cluster undergoes a 10:3:1 
ratio triple merger accompanied by a dramatic rise in its integrated Compton-Y, peaking at 
$z = 0.05$. By modeling the thermal SZE (tSZ) and kinetic SZE (kSZ) spectral distortions of the 
Cosmic Microwave Background (CMB) at this redshift with relativistic corrections, we produce 
various mock images of the cluster at frequencies and resolutions achievable with current 
high-resolution SZE instruments. The two gravitationally-bound merging subclusters account 
for 10\% and 1\% of the main cluster's integrated Compton-Y, and have extended merger 
shock features in the background ICM visible in our mock images. We show that along 
certain projections and at specific frequencies, the kSZ CMB 
intensity distortion can dominate over the tSZ due to the large line of sight velocities of 
the subcluster gas and the unique frequency-dependence of these effects. We estimate that 
a one-velocity assumption in estimation of line of sight velocities of the merging subclusters 
from the kSZ induces a bias of $\sim$10\%. This velocity bias is small relative to other sources 
of uncertainty in observations, partially due to helpful bulk motions in the background ICM 
induced by the merger. Our results show that high-resolution SZE observations, which have 
recently detected strong kSZ signals in subclusters of merging systems, can robustly probe 
the dynamical as well as the thermal state of the ICM.
\end{abstract}


\section{Introduction}

	Galaxy clusters are the most massive gravitationally-bound objects in the
Universe, and their abundance evolution provides a direct test of the growth of structure. 
Cluster surveys have provided constraints on cosmological parameters such as 
$\Omega_M$, $\Omega_\Lambda$, $\sigma_8$, and $w_0$ \citep[e.g.][for a recent review,
see Allen et al. 2011]{ma08, ma10, vi09, ro10, va10, se11}. While cluster research has
traditionally relied heavily on X-ray observations, new survey telescopes such as the 
\emph{Planck} satellite \citep{pl11a}, the Atacama Cosmology Telescope \citep[ACT,][]{fo07}, 
and the South Pole Telescope \citep[SPT,][]{ca11} are instead probing the Sunyaev-Zel'dovich 
Effect \citep[SZE,][]{ze69, su70, su72}. Observations of the SZE detect small distortions in the 
Cosmic Microwave Background (CMB) blackbody spectrum caused by inverse-Compton 
scattering of CMB photons off energetic electrons in the hot intracluster medium (ICM) 
\citep[for reviews, see][]{bi99, ca02}. Since this spectral distortion does not suffer from 
cosmological surface brightness dimming, the SZE has the advantage of being able to detect
clusters out to higher redshifts than possible in the X-ray. 

	Current blind SZE surveys scan large regions of the millimeter and sub-mm sky to 
detect intensity fluctuations in the CMB from the SZE of individual clusters. These observations
have relatively low angular resolutions ($\gtrsim$1\arcmin) and are typically unable to resolve 
cluster substructures, except in exceptional cases of massive nearby clusters such as the Coma cluster \citep{pl12a}. Thus, the resulting key observable is the total Compton-Y, integrated 
over the entire cluster (Equation 1). Using well-known cluster scaling relations between 
various observed quantities such as Y and the total cluster mass M \citep[e.g.][]{ar10, ma12, 
pl12b, si12, ho12}, masses for large samples of clusters can be directly inferred. While the Y-M 
scaling relation is tight, there is some intrinsic scatter which may be induced by various 
phenomena such as cluster mergers and AGN \citep[e.g.][see however Poole et al. 2007]{kr12, ba12}. A clear 
understanding of the sources of intrinsic scatter in cluster scaling relations is crucial, as this scatter 
will limit the precision of cosmological parameters inferred from SZE surveys. 

	The advent of high-resolution (10-30\arcsec) SZE imaging of clusters opens a new 
window to the thermal and dynamical state of the ICM. These observations have the ability to
resolve small-scale disturbances in the ICM caused by various phenomena, thus
probing their effects on the total Compton-Y of each cluster. Bolometer arrays such as the 
Multiplexed SQUID/TES Array at Ninety GHz \cite[MUSTANG,][]{di08} on the Green Bank 
Telescope (GBT) and Bolocam \citep{ha04} on the Caltech Submillimeter Observatory 
(CSO), as well as interferometric dish arrays such as the Combined Array for Research in 
Millimeter-wave Astronomy \citep[CARMA,][]{bo06} and the Atacama Large Millimeter/submillimeter 
Array (ALMA) are beginning to make these unique observations at a variety of frequencies  
\citep{ki04, no09, mas10, ko11, plag12, re12, mr12}.

	In this paper, we focus specifically on the impact of cluster mergers in high-resolution 
SZE observations. This is particularly relevant as clusters are assembled through hierarchical 
growth via mergers in the standard $\Lambda$CDM cosmology \citep[][and references 
therein]{kra12}. Since blind SZE cluster surveys will probe high redshifts, a significant fraction 
of SZE-selected clusters will still be experiencing ongoing mergers. Hydrodynamic simulations 
of cluster mergers have revealed transient boosts in the total Compton-Y during passages
of merging subclusters through the main cluster \citep{po07, wi08}. 
This can cause SZE surveys to preferentially detect merging systems with overestimated cluster 
masses, thus biasing the inferred cosmology to higher $\sigma_8$, and lower $\Omega_M$ 
\citep{ra02, wi08}. Indeed, X-ray follow-up of clusters detected by \emph{Planck} have unveiled 
a new and significant population of disturbed systems \citep{pl11b, pl12c}. Clusters appearing 
to be relaxed in X-ray observations have also been found to contain substructures attributed to 
mergers in high-resolution SZE observations. In the case of RXJ1347.5-1145, substructures in 
10$\arcsec$ resolution SZE observations interpreted as a merger shock are estimated to be a 
10\% effect on observations with 1$\arcmin$ resolution \citep{mas10, ko11}. 

	High-resolution SZE imaging of more clusters will be crucial for understanding the 
effects of mergers on the integrated Compton-Y, and important for precision cluster cosmology.
These observations will require guidance from hydrodynamic cluster simulations to provide a 
framework for proper interpretation. For example, observed SZE substructures in merging systems 
are combinations of thermal SZE (tSZ) and kinetic SZE (kSZ) contributions from merger shocks, 
the remnant cores of merging subclusters, and the ambient background ICM; simulations are 
necessary to identify means of disentangle these different effects. Previous studies of the kSZ
using cosmological simulations have often focused on the peculiar motions of unresolved individual 
galaxy clusters as a whole \citep[e.g.][]{ro07}, or relatively relaxed (non-merging) clusters \citep[e.g.][]{di05}, 
although \citet{ma07} has investigated the kSZ polarization in merging systems. It is not yet clear 
how the observable tSZ and kSZ morphology of a cluster will change during a merger, especially 
at different projections. Our main goal in this paper is to show what a typical merging cluster detected 
in blind SZE surveys will look like in follow-up high-resolution SZE observations, and how the tSZ 
and kSZ contribute to the resulting SZE signal. Using a cosmological simulation of a massive cluster 
undergoing a 10:3:1 triple merger described in Section 2, we model its frequency-dependent 
tSZ and kSZ distortion on the CMB spectrum. In Section 3, we produce mock SZE intensity maps 
of the merging cluster at different frequencies and resolutions achievable by current instruments. 
In Section 4, we discuss the observable SZE substructures from the merger shocks and the 
kSZ contributions to the observable SZE signal. To separate the sources of the observable 
SZE signal, we decompose the cluster into its gravitationally-bound components to probe the 
separate SZE contributions of individual substructures and the background ICM (including the 
merger shocks). We show that the kSZ can dominate the CMB spectral distortion for merging 
subclusters at certain projections, and find that multi-frequency SZE-inferred velocities of merging 
subclusters are biased to only $\sim$10\% from the one-velocity approximation partially due to 
additional motions in the background ICM induced by the merger. We summarize our results 
and conclude in Section 5.

	Throughout the paper, we adopt a standard $\Lambda$CDM cosmology with 
$\Omega_M = 0.28$, $\Omega_\Lambda = 0.72$, $\Omega_b = 0.046$, $\sigma_8 = 0.82$, 
and $H_0 = 70$ km s$^{-1}$ Mpc$^{-1}$, consistent with the 5 year \emph{WMAP} results of \citet{ko09}.

\section{The Simulated Cluster}
	Our cluster was simulated using the cosmological TreeSPH code 
ChaNGa \citep{gi07} and the zoomed-in technique of \citet{ka93}. An initial dark matter 
(DM) only simulation of a 400$h^{-1}$ Mpc periodic cube with $2.16\times10^8$ 
DM particles of mass $3.29\times10^{10}$ M$_{\odot}$ was run with initial mass perturbations
generated using CMBFAST \citep{se96} and the above cosmological
parameters. This low-resolution simulation was run to $z = 0$, and
all resulting cluster haloes were identified using the friends-of-friends algorithm with a
linking length of 0.2 times the mean interparticle separation.
Formation times were determined for each cluster by determining the
epoch at which the most massive progenitor reached 75\% of its $z = 0$
mass. We chose a cluster to resimulate whose formation time was
approximately the median of all clusters greater than $1.5 \times 10^{15}$ M$_{\odot}$ 
(note that this mass defined by the friends-of-friends algorithm is slightly different from 
the virial mass we use below). The initial conditions for the subsequent high-resolution
simulation were generated by identifying the Lagrangian region in the
lower-resolution simulation that ends up within 3 virial radii of the
chosen cluster, and resolving this region with DM particles of mass $5.37
\times 10^{7}$ M$_{\odot}$ and gas particles of mass $1.06\times 10^{7}$
M$_{\odot}$. This region is surrounded by layers of lower-resolution
spherical shells out to the full 400$h^{-1}$ Mpc box. Additional 
high-resolution perturbations were also applied to particles in the central
64$h^{-1}$ Mpc of this box. These initial conditions for the zoomed-in simulation
were then evolved from $z = 139$ to the present with an `adiabiatic' equation of
state for the gas. That is, the gas is evolved with no radiative cooling, non-gravitational 
heating, or star formation, but the gas evolution is not isentropic due to the presence of shocks.
Inside the virial radius of this cluster at $z = 0.05$ (the output at which our SZE analysis 
is performed), there are only five low-resolution dark matter particles, leading to a total mass 
of $\sim$$8 \times 10^{8}$ M$_{\odot}$ in `missing' gas particles. All the low-resolution 
DM particles inside the cluster are near the virial radius and  are not associated with any 
subhaloes of the main cluster. This confirms that our simulation analysis is robust to our choice of 
the size of the zoomed-in region. All analysis and visualization was done using the 
parallelized simulation visualization tool Salsa\footnote{http://software.astro.washington.edu/nchilada/}, 
unless otherwise stated.

\begin{figure}
\centering
\includegraphics[width=0.47\textwidth]{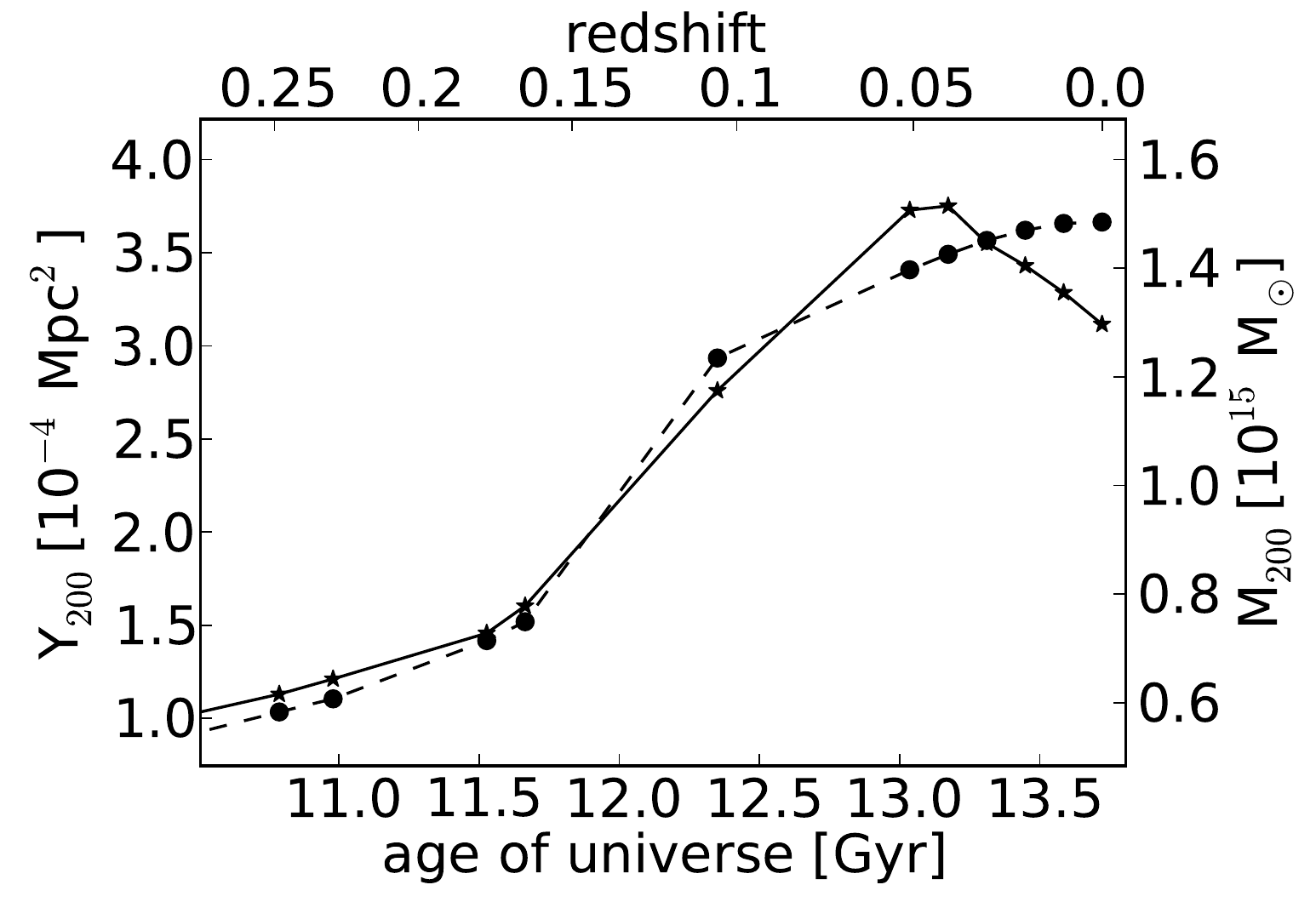}
\caption{
The time evolution of the total integrated Y$_{200}$ (solid) and M$_{200}$ (dashed) of the 
main cluster. The epochs sampled by our simulation outputs are also shown.
A 10:3:1 triple merger occuring between $z \sim 0.17$ to 0.12 causes a dramatic rise in 
M$_{200}$. There is a corresponding rise in Y$_{200}$, although its evolution after the merger 
deviates from that of M$_{200}$ as the cluster relaxes.
}
\label{fig:SZevol}
\end{figure} 

	We identify the cluster centre as the particle at the deepest potential, and determine the
properties of the cluster in the spherical volume with an overdensity of 200$\rho_{cr}$ 
(where $\rho_{cr}$ is the critical density of the Universe). At $z=0$, the virial 
radius  is R$_{200}$ = 2.35 Mpc, and the total virial mass is M$_{200, tot} = 1.48\times10^{15}$ 
M$_{\odot}$. This total mass includes a gas mass of M$_{200, gas} = 2.22\times10^{14}$ 
M$_{\odot}$ and a DM mass of M$_{200, DM} = 1.26\times10^{15}$ M$_{\odot}$. 
Figure~\ref{fig:SZevol} shows the evolution
of M$_{200, tot}$ of this cluster over time. The cluster slowly accretes mass from $z \sim 0.28$ 
to 0.17, but then undergoes a period of dramatic mass gain due to a triple merger from 
$z \sim 0.17$ to 0.12. The mass increase from the triple merger is not sharp in 
Figure~\ref{fig:SZevol} due to the limited time resolution of the outputs. After the mergers, 
the mass accretion decreases to rates comparable to that from before the merger. By visualizing the 
simulation at $z = 0.23$ (before the mergers), we identify the three pre-merger clusters to 
determine that the triple merger is approximately 10:3:1 ratio in mass. 

\begin{figure}
\centering
\includegraphics[width=0.47\textwidth]{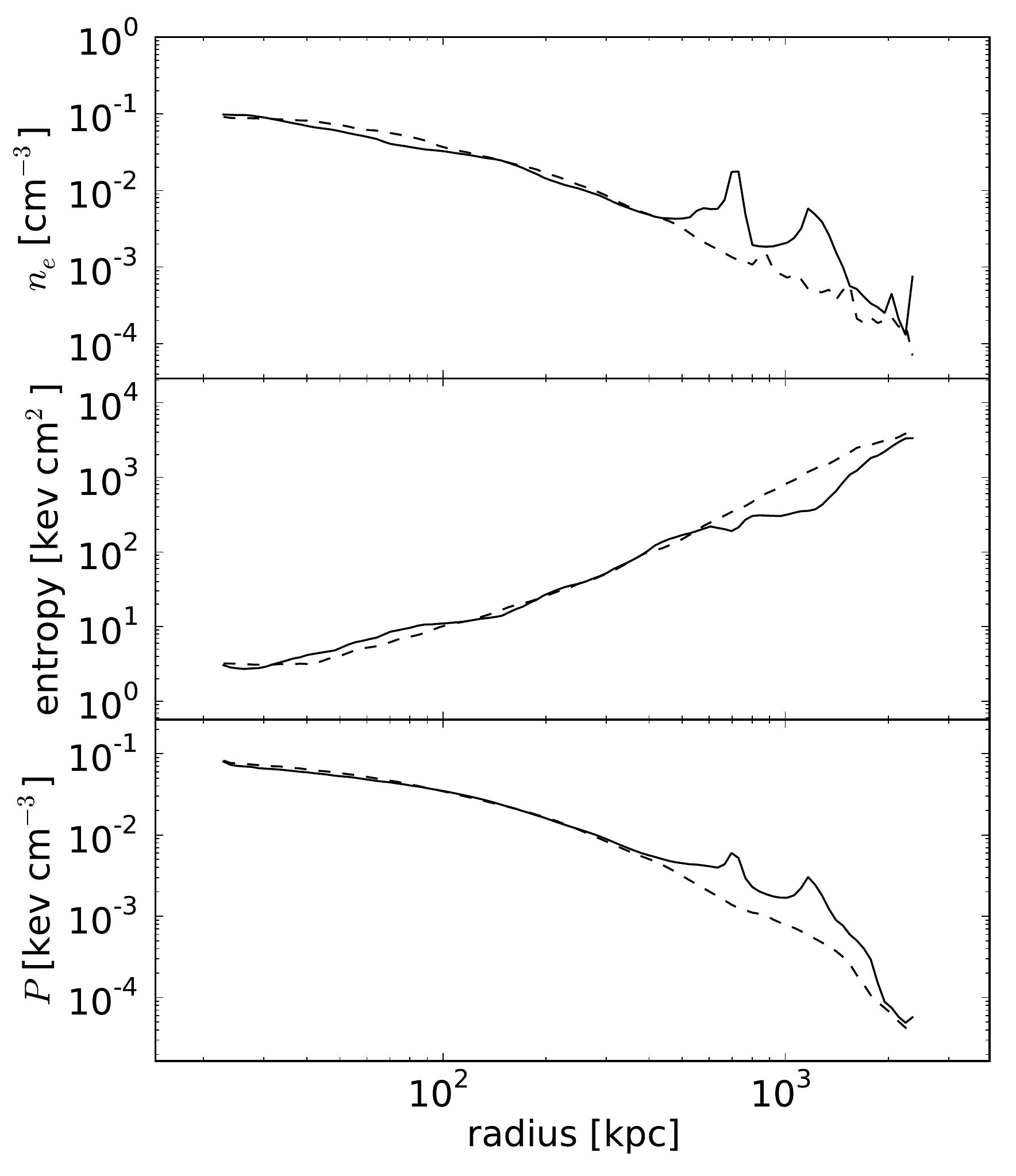}
\caption{
Radial profiles of electron number density (top), entropy (middle), and gas pressure (bottom), of
the quadrant of the main cluster containing substructures A and B (solid), and the remaining three
quadrants for comparison (dashed). The profiles extend out to R$_{200}$, and are of the cluster at output 
$z = 0.05$. The effects of the merging substructures A (at 1.14 Mpc) and B (at 0.7 Mpc) are evident.
Due to the non-radial trajectories of these merging substructures, the entropy jumps due to 
their merger shocks are not clearly visible in these radial profiles.
}
\label{fig:profiles}
\end{figure}

	The observable tSZ can be parameterized using the total Compton-Y parameter, which is
integrated over the full angular extent of the cluster. In our simulation, this is calculated over all 
gas particles inside R$_{200}$ of the cluster at each time step using
\begin{equation}
	Y_{200} = \int_{V_{200}} n_e \sigma_T \frac{k_\mathrm{B} T}{m_e c^2}\,\mathrm{d}V.
\end{equation} 
Due to the relatively low densities and temperatures of the gas outside R$_{200}$, the observable 
SZE signal beyond R$_{200}$ is negligible. The time evolution of Y$_{200}$ is also shown in 
Figure~\ref{fig:SZevol}. The Y$_{200}$ increases slowly before $z \sim 0.17$, and also 
experiences a dramatic rise due to the triple merger.
As the mass accretion slows down after the merger, Y$_{200}$ continues 
to rise from $z \sim 0.12$ to 0.05, likely due to shocks that form as the 
merging subclusters travel through the denser interior of the main cluster. Visualization of 
the cluster at $z = 0.05$ shows that the merging subclusters in the 10:3:1 merger are in the late 
stages of their initial pass through the main cluster, with well-developed merger shocks. However, 
after peaking at  $z \sim 0.05$, the Y$_{200}$ then sharply decreases as the cluster 
begins to relax. Transient boosts in Y$_{200}$ are also seen in the idealized merger 
simulations of \citet{po07}. The evolutionary tracks of simulated clusters along the Y-M relation 
during mergers have been investigated in previous studies \citep[e.g.][]{po07, kr12}, and these 
deviations are a known source of intrinsic scatter in Y-M scaling relations (but may not be 
able to account for all of the scatter). While we do not focus on the exact evolution of the 
Y-M relation in this paper, the peak of the Compton-Y at $z \sim 0.05$ suggests that the 
SZE substructures associated with the merger is likely to have the biggest effect on Y$_{200}$ at 
this redshift. Furthermore, this peak of the Compton-Y is also when this cluster is most 
detectable in blind SZE surveys, and so we focus our simulation analysis on this output. 

	Figure~\ref{fig:profiles} shows radial profiles of the mass-weighted electron number 
density $n_e$, gas entropy $k_\mathrm{B} n_e^{-2/3} T$, and gas pressure $P$ at 
$z = 0.05$. We calculate profiles for the quadrant of the cluster containing both merging 
subclusters, as well as the remaining three quadrants for comparison. The pressure profile
for the three quadrants without merging subclusters are in broad agreement with observations
of the most massive clusters in \citet{ar10}, as well as the Coma cluster \citep{pl12a}.
At this epoch, the larger merging subcluster in the 10:3:1 triple 
merger (hereafter substructure A) is 1.14 Mpc from the center of the main cluster, and 
the cold, dense gas in its remnant core is clearly seen as a jump in the average $n_e$ and $P$ 
at this radius. The strong contrast in $P$ between substructure A and the background ICM should 
make substructure A clearly visible in high-resolution SZE images. The smaller merging subcluster 
in the triple merger (hereafter substructure B) is also visible as a second peak in the $n_e$ and 
$P$ profiles at a radius of 0.70 Mpc. Visualization of this cluster clearly shows merger shocks 
in front of both substructures A and B in the form of strong temperature increases. However, 
sharp increases in the entropy profile from the shocks around 0.07 and 1.14 Mpc are not seen
in Figure~\ref{fig:profiles} because neither merging substructures have trajectories aligned with 
the center of the main cluster (i.e., a large component of the velocity vectors of both shocks are in the 
tangential direction). This purely geometrical effect spreads out the entropy jump from the shock 
over a large range of radii in Figure~\ref{fig:profiles}. In fact, the entropy profile actually decreases 
around both substructures due to the low entropy of the cold gas in their remnant cores. 

\section{Mock Sunyaev-Zel'dovich Effect Images}

\begin{figure*}
\centering
\includegraphics[width=0.49\textwidth]{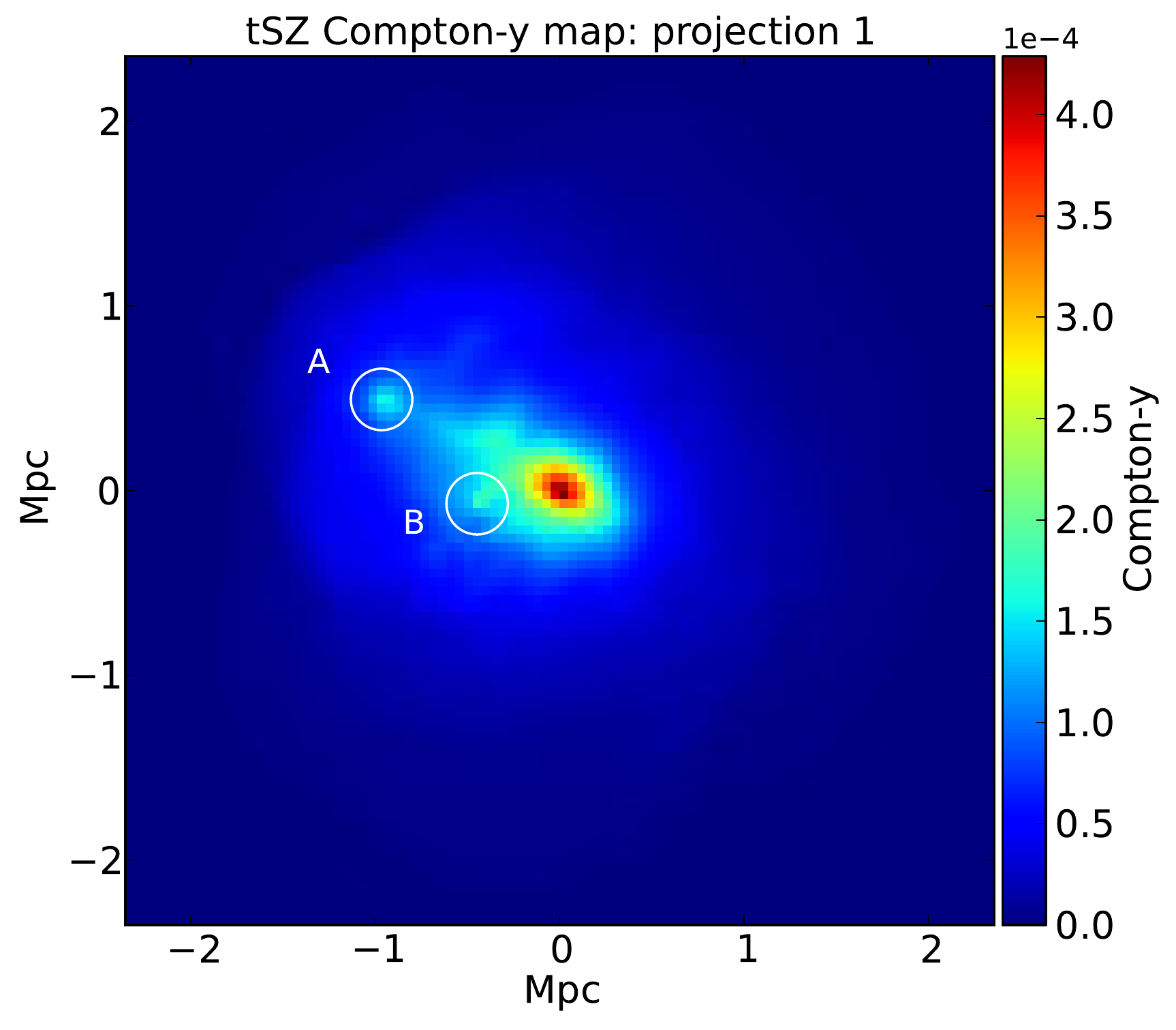}
\includegraphics[width=0.49\textwidth]{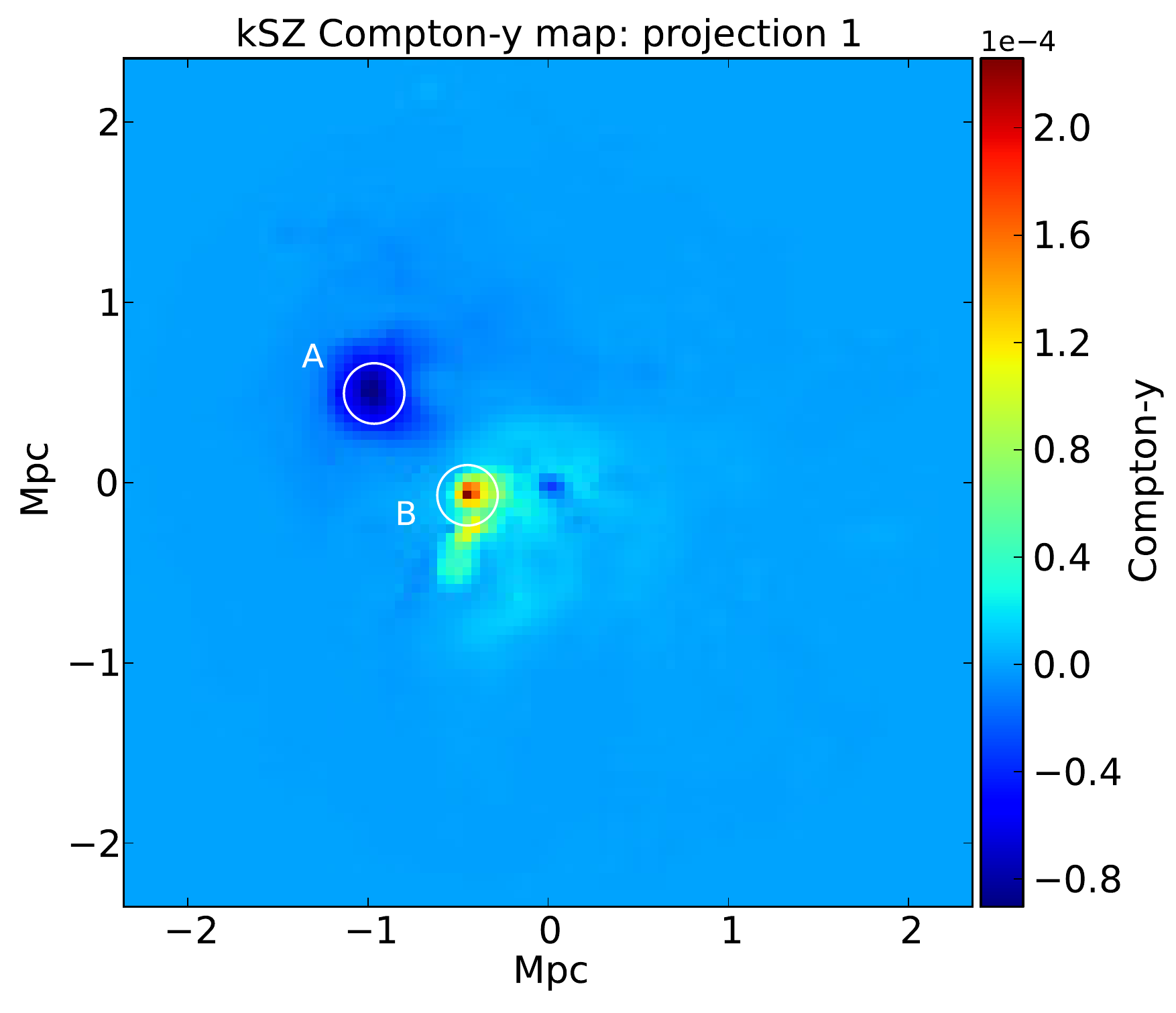}\\
\includegraphics[width=0.49\textwidth]{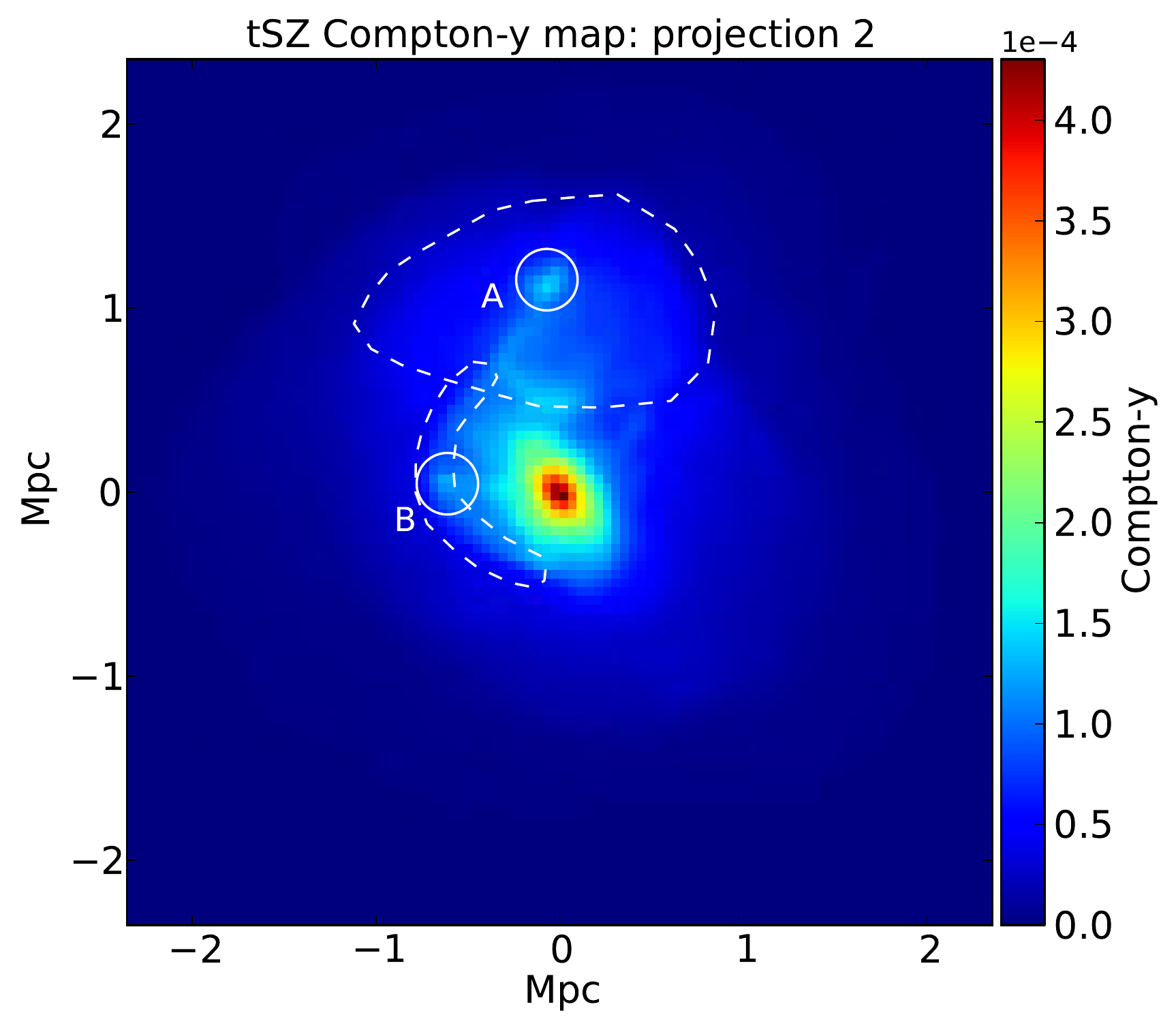}
\includegraphics[width=0.49\textwidth]{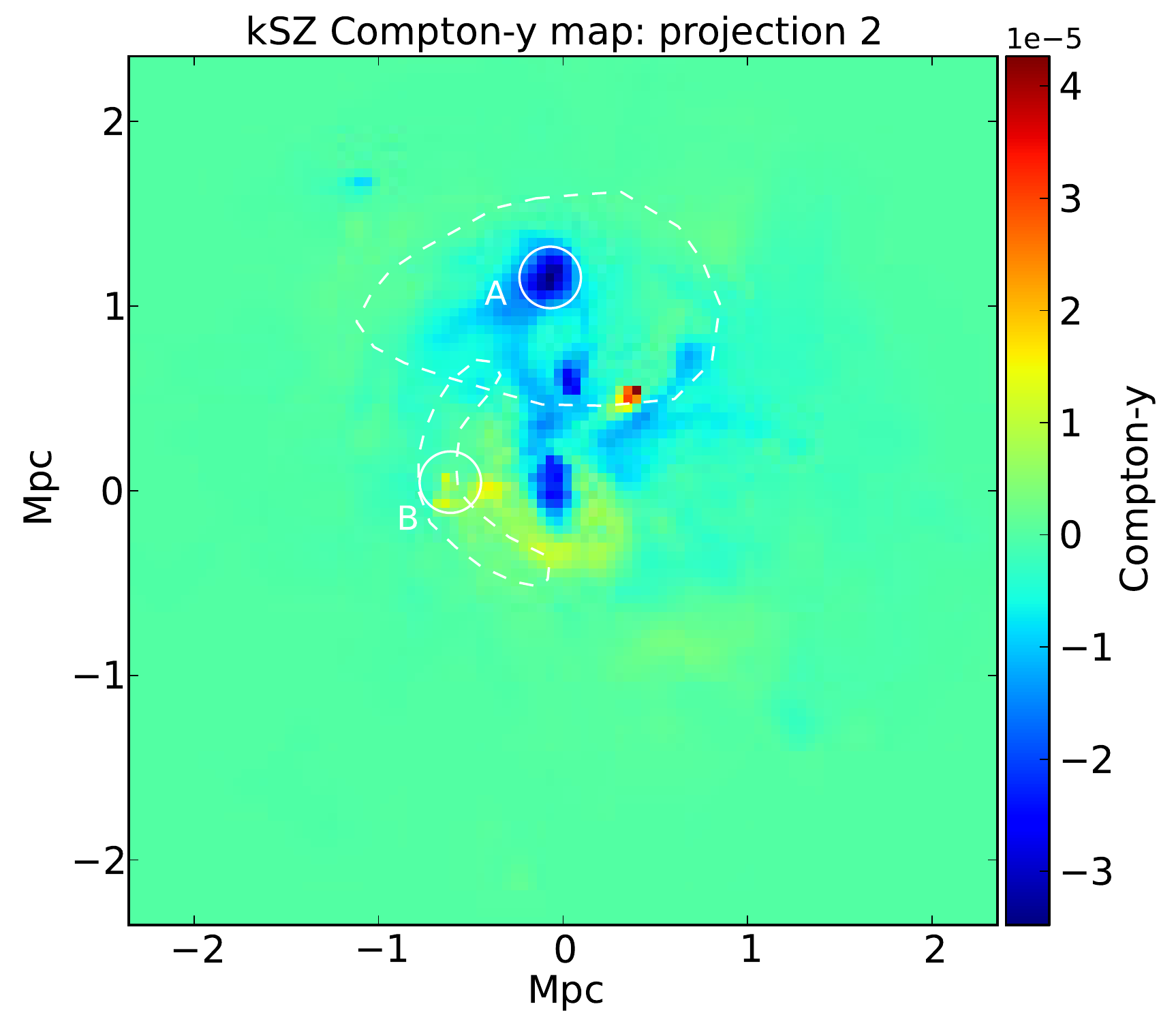}
\caption{
Maps of the average Compton-y of the gas along the line of sight of each pixel of the cluster, 
due to the tSZ (left panels, see Equation 2) and kSZ (right panels, see Equation 3), along 
projections 1 (top panels) and 2 
(bottom panels). Merging substructures A and B are encircled. These two projections are orthogonal, 
and both substructures have significant (and opposite) line of sight velocities in projection 1, 
which are transverse velocities in projection 2. The merging substructures are clearly
resolved in the kSZ Compton-y map of projection 1 (top right). In the tSZ Compton-y map of 
projection 2 (bottom left), a bow shock surrounding substructure B (traveling leftwards) and 
faint extended regions of shocked gas around substructure A (traveling upwards) are visible and encircled
with dashed lines. Each pixel is $50\arcsec\times50\arcsec$ (corresponding to 47 kpc $\times$ 
47 kpc) if the cluster is at a distance corresponding to a redshift of $z = 0.05$.
}
\label{fig:SZcomptony}
\end{figure*}

	We produce projected tSZ and kSZ maps of the cluster's R$_{200}$ spherical volume at 
$z = 0.05$ by binning the field of view of R$_{200}$ $\times$ R$_{200}$ along two orthogonal 
projections into 100 $\times$ 100 pixels, each 47 kpc $\times$ 47 kpc physically. Projection 1
was selected such that the line of sight velocities of the merging subclusters 
in the cluster rest-frame are maximal, while projection 2 is in the perpendicular direction.
We expect any interesting effects from the kSZ to be strongest in projection 1, and 
weakest in projection 2. However, the strong transverse velocities of the substructure 
gas in the second projection allows us to study the large-scale effects of the merger shocks on the 
main cluster's ICM. Figure~\ref{fig:SZcomptony} shows projected maps of the cluster's tSZ 
Compton-y (left panels), mass-weighted over all gas particles inside the cluster virial radius in 
each pixel, and calculated using 

\begin{equation}
	y_{tSZ} = \int n_e \sigma_T \frac{k_\mathrm{B} T}{m_e c^2}\,\mathrm{d}l,
\end{equation} 
as well as the kSZ Compton-y in the cluster rest-frame (right panels),
\begin{equation}
	y_{kSZ} = \int  n_e \sigma_T \frac{v}{c}\,\mathrm{d}l.
\end{equation} 

	We compute the kSZ in the rest-frame of the main cluster. In general, the observed
kSZ signal will also include a small contribution from the peculiar motion of the cluster itself 
\citep{ho97, be03, ha12, mau12, ze12}. Our simulated cluster has a bulk gas velocity of
159 km s$^{-1}$, consistent with constraints from observations \citep{pl13}. More importantly,
in projection 1 (the projection selected to optimize the kSZ signal from the merging subclusters),
the main cluster's bulk line of sight velocity is only 33 km s$^{-1}$, much smaller than the velocities 
of the merging subclusters (Sections 4.1 and 4.2). For this reason, we do all our analysis in the 
rest-frame of the main cluster, which does not significantly affect our results, and allows for easy 
comparison to other simulations as well as observations.

	Substructures A and B are both visible as pressure enhancements in the tSZ 
Compton-y maps shown in Figure~\ref{fig:SZcomptony}. The main cluster is clearly unrelaxed 
and asymmetric, as there are large-scale disturbances in the ICM around both substructures. 
The kSZ Compton-y map of projection 1 (top right panel of Figure~\ref{fig:SZcomptony}) 
reveals strong and opposing line of sight motions for substructures A and B. 
We discuss the origins and implications of all these substructures in Section 4.
We considered extending the integration depth in each pixel for 
Equation 2 to 4 $\times$ R$_{200}$, but because the temperature and density of the 
gas outside the virial radius is so low, the average Compton-y in each pixel did not change 
by more than $\sim$1\%.

\begin{figure*}
\centering
\includegraphics[width=0.98\textwidth]{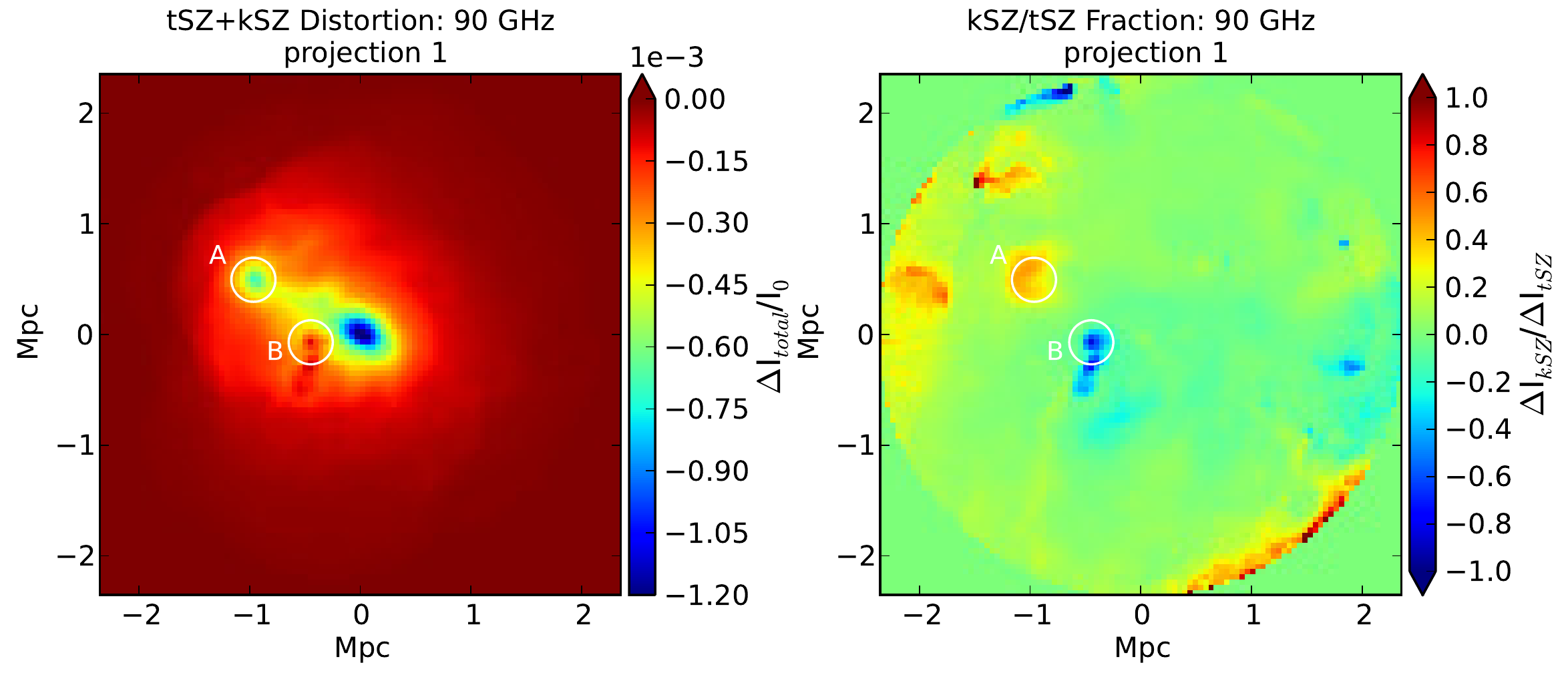} \\
\includegraphics[width=0.98\textwidth]{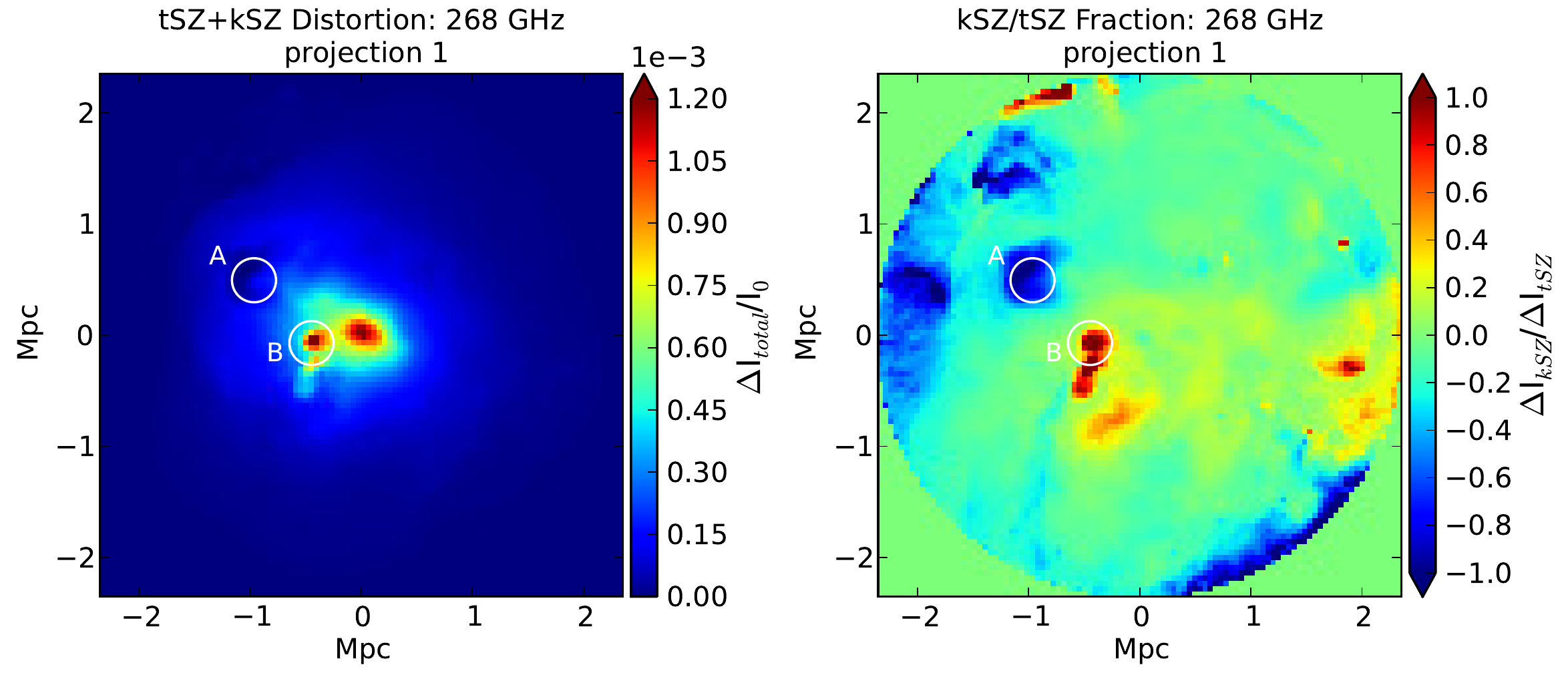}
\caption{
Mock images of the total tSZ+kSZ intensity distortion (see Equations 4 and 6) at 90 GHz and 268 GHz
in projection 1 (left panels). The appearance and disappearance of substructures 
A and B at these different frequencies in projection 1 is due to unique frequency-dependence of 
the kSZ, which can dominate the spectral distortion along merging substructures. 
Maps of the kSZ/tSZ intensity distortion ratio in these images are also shown (right panels). 
Large-scale features in these kSZ/tSZ ratio maps not associated with either substructure 
are due to bulk gas motions in low-density gas, and are not observable in the left panels. 
}
\label{fig:SZdistortion1}
\end{figure*}

\begin{figure*}
\centering
\includegraphics[width=0.98\textwidth]{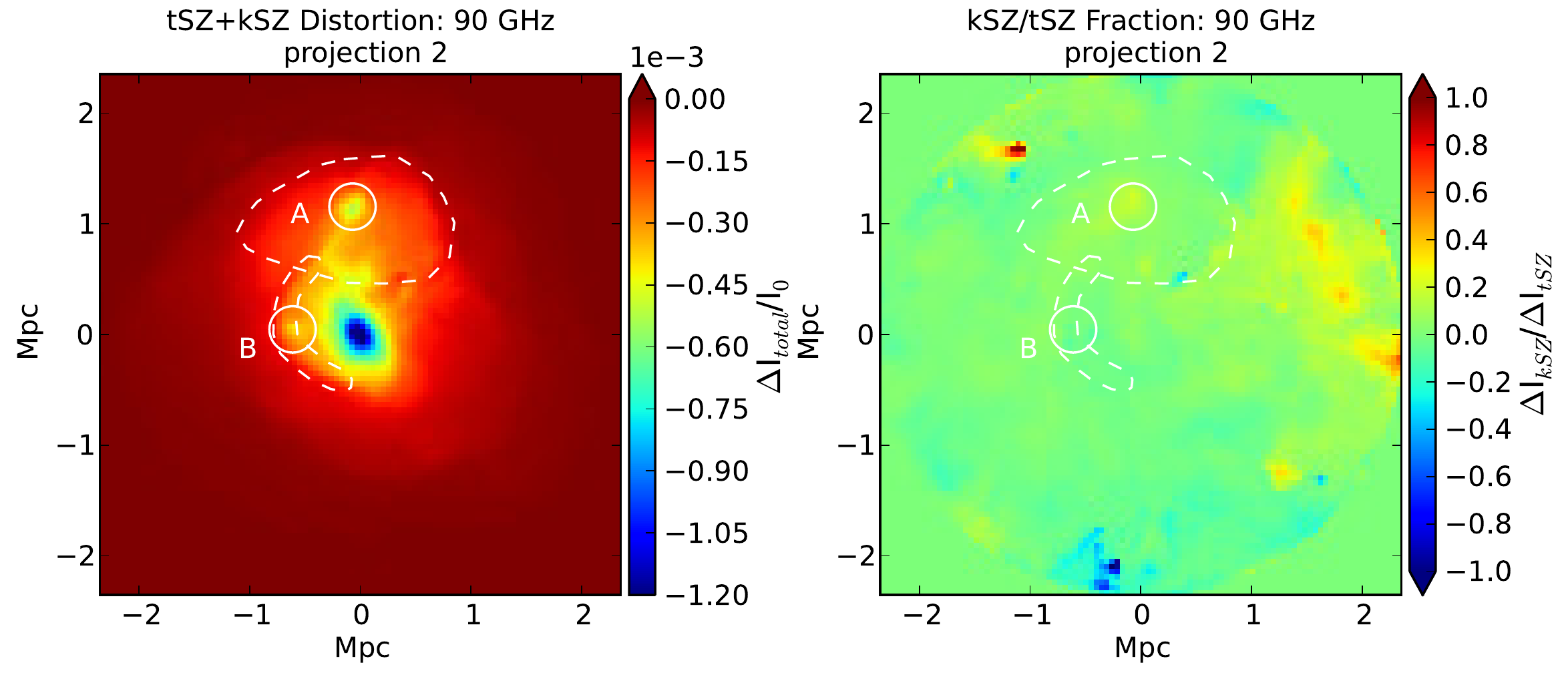} \\
\includegraphics[width=0.98\textwidth]{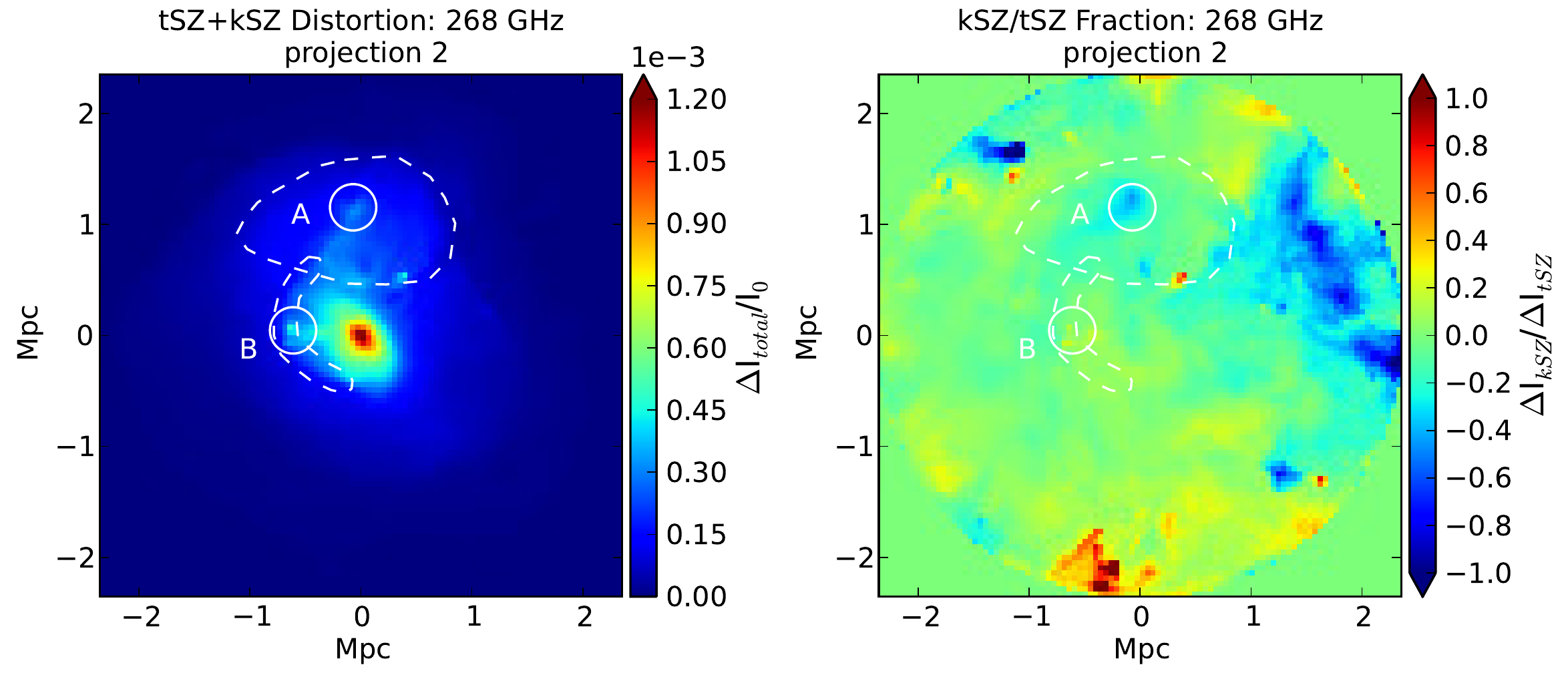}
\caption{
Mock images of the total tSZ+kSZ intensity distortion similar to Figure~\ref{fig:SZdistortion1}, but in 
projection 2. Shock features, including a bow shock around substructure B (traveling leftwards) 
and extended regions of shocked gas around substructure A (traveling upwards) are visible
and encircled with dashed lines.
}
\label{fig:SZdistortion2}
\end{figure*}

	Observationally, the pixels in Figure~\ref{fig:SZcomptony} are $10\arcsec\times10\arcsec$ 
if the cluster is located at a distance corresponding to $z = 0.32$, or $50\arcsec\times50\arcsec$ 
if the cluster is at distance $z = 0.05$. While this sub-arcminute resolution is achievable by current 
single-dish high-resolution SZE instruments such as MUSTANG (with resolution of 
$\sim$10$\arcsec$), the filtering techniques used to remove atmospheric noise in real MUSTANG 
SZE maps are unable to recover signals on scales larger than $\sim$45$\arcsec$. Interferometric 
dish arrays have larger spatial dynamic ranges, and arrays such as CARMA are able to probe 
angular scales of $\sim$$15\arcsec$-$5\arcmin$. While the full angular size of our mock SZE maps 
may not possible for large clusters at low redshifts, the results of our investigation are also relevant 
for mergers occurring closer to the core of the main cluster and SZE-selected clusters at higher 
redshifts, where angular dynamic range issues are less important. Future instruments and 
improvements in observational techniques in this emergent field will also be able to increase the 
angular dynamic range, and so our work also serves as an exploratory proof-of-concept for the 
science possibilities. 

	 The observable frequency-dependent tSZ spectral distortion of the CMB can 
be calculated as the change in specific intensity from the Planckian CMB,
\begin{equation}
\Delta I_{tSZ} = I_0  y_{tSZ} \frac{x^4 e^x}{(x^2 -1 )^2}(x \coth \left(\frac{x}{2}\right)-4) g_{tSZ}(x)
\end{equation}
at dimensionless frequency
\begin{equation}
x = \frac{h \nu}{k_\mathrm{B} T},
\end{equation}
where  $I_0$ is the specific intensity of the CMB, and
$g_{tSZ}(x)$ are the relativistic corrections calculated in \citet{no98}. The frequency-dependent 
spectral distortion due to the kSZ can similarly be calculated using
\begin{equation}
\Delta I_{kSZ} = I_0  y_{kSZ} \frac{x^4 e^x}{(x^2 -1 )^2} g_{kSZ}(x).
\end{equation}
Figures~\ref{fig:SZdistortion1} and \ref{fig:SZdistortion2} shows mock images of the CMB 
intensity distortion due to the SZE in projections 1 and 2. We chose to generate
these mock images at 90 GHz and 268 GHz, frequencies probed by instruments such as 
MUSTANG on GBT and Bolocam on CSO, respectively. We also show the kSZ/tSZ 
intensity distortion fraction in each pixel, and discuss its implications in Section 4.1.

\section{The Effects of Mergers}

\subsection{SZE Contributions from the kSZ}

	While most high-resolution SZE observations thus far have been done in a single frequency
band, recent progress have been made to image clusters at multiple frequencies by combining 
observations from different instruments. This multi-frequency approach is crucial
to break the degeneracy between the tSZ and kSZ that exist in single-band observations, 
by leveraging the unique frequency-dependence of these two effects (see Equations 4 and 6).
Recently, \citet{mr12} have shown that significant contributions from the kSZ is necessary
to model multi-frequency SZE observations of substructures in a merging system. Based on the
inferred kSZ, \citet{mr12} calculated a line of sight velocity of the substructure that is consistent with
those determined from optical spectroscopy of the galaxies. Although this is a tantalizing result,
these observations carry some uncertainties in the measurement and modeling, as well as the 
precise origin of the SZE substructure. Using our ability to model the SZE in our cluster simulation 
and separate the merging substructures from the ICM, we can investigate the kSZ contribution to the 
observed SZE and the origins of these observed substructures.

\begin{figure}
\centering
\includegraphics[width=0.45\textwidth]{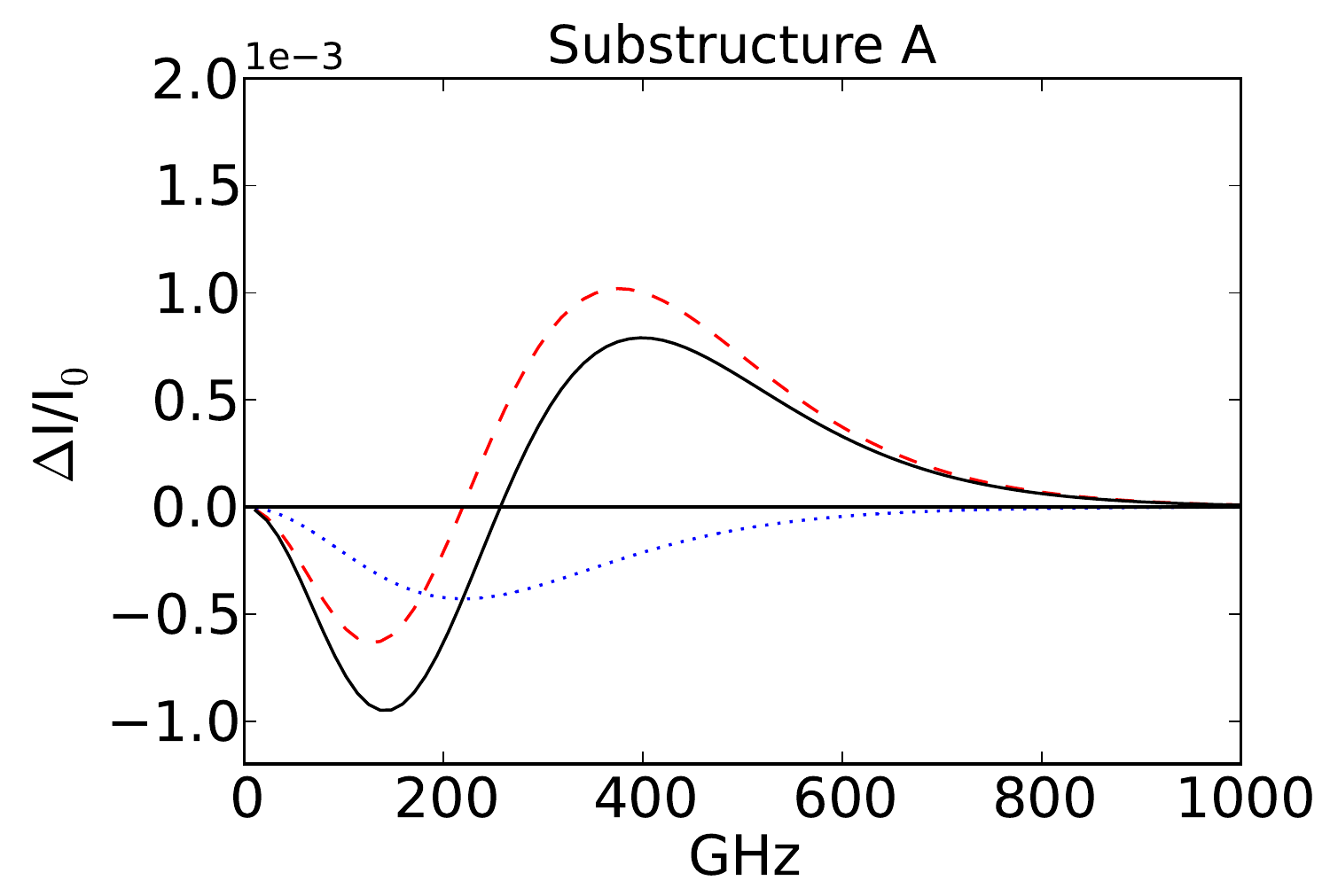}\\
\includegraphics[width=0.45\textwidth]{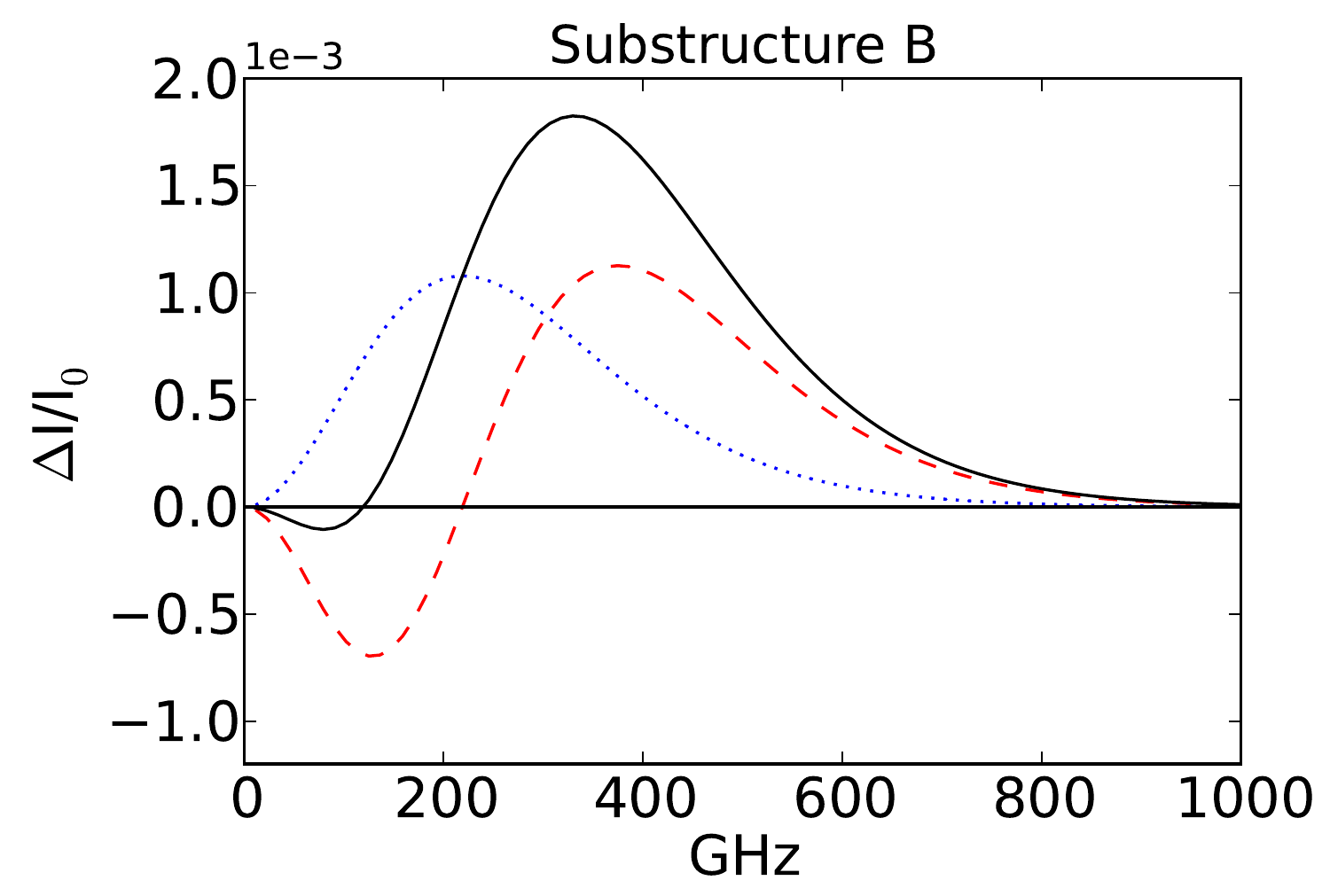}
\caption{
Relativistically-corrected CMB spectral distortions of the gas along the line of sight of the 
remnant cores of substructure A (top) and B (bottom) in projection 1, due to the tSZ (red dashed), 
kSZ (blue dotted), and the combined t+kSZ distortion (black solid). 
}
\label{fig:SZspectrum}
\end{figure}

	The mock CMB spectral distortion maps of Figures~\ref{fig:SZdistortion1} and \ref{fig:SZdistortion2} 
includes distortion from both the tSZ and kSZ, taking into account their frequency dependence. 
Projection 1 of our mock SZE maps was selected to highlight the kSZ since in this projection,
substructure A has line of sight velocity of $-1150$ km s$^{-1}$ (away from the observer),
while substructure B has line of sight velocity of $2492$ km s$^{-1}$ (towards the observer) in
the rest-frame of the cluster. We see interesting effects in Figure~\ref{fig:SZdistortion1} around the 
substructures when observations at different frequencies are possible. In particular, substructure 
A appears as a strong CMB intensity decrement at 90 GHz, but is absent in the mock image at 
268 GHz. This effect also occurs for substructure B, but in the opposite sense. This can be 
understood as the effects of the unique frequency-dependence of the tSZ and kSZ, which will cause 
their relative contributions to the total CMB distortion to vary at different frequencies. 
These effects are illustrated in Figure~\ref{fig:SZspectrum}, which shows the relativistically-corrected CMB 
spectral distortion for substructures A and B. These spectra have been modeled using gas along the 
line of sight of the pixel in projection 1 that corresponds to the dense remnant core of each 
substructure. We can see that for substructure A, its negative line of sight velocity causes a
kSZ decrement at 90 GHz, which combines with the negative tSZ decrement at this frequency to 
produce a large total CMB intensity decrement. At 268 GHz (on the other side of the 217 GHz tSZ null),
the tSZ distortion becomes an intensity increment while the kSZ distortion is still a decrement,
and so the combination of the two conspires to nullify the SZE distortion. These frequency-dependent 
effects occur similarly for substructure B, but at opposite frequencies since its
positive line of sight velocity causes the kSZ distortion to become an increment instead at all frequencies.

	In Figures~\ref{fig:SZdistortion1} and \ref{fig:SZdistortion2} (right panels), we have 
also shown maps of the kSZ/tSZ CMB intensity distortion fraction. Along the line of sight of 
substructures A and B in projection 1, the kSZ can contribute significantly to the total SZE 
distortion, and even dominate over the tSZ. This is extremely important for interpreting 
high-resolution SZE observations of merging systems, as misinterpretations of the source of 
SZE substructure may occur if the tSZ-kSZ degeneracy cannot be broken. Single-frequency 
observations of SZE substructures have often assumed that the kSZ is negligible, which we have 
shown may not be true in all cases, and so proper interpretation will require multi-frequency 
observations. Our results showing significant kSZ contributions in merging subcusters should
not be surprising, as these effects have already been tentatively observed in \citet{mr12}. 
We note that regions with large kSZ/tSZ CMB intensity distortion fraction in 
Figures~\ref{fig:SZdistortion1} and \ref{fig:SZdistortion2} (right panels) not associated with 
substructures A and B are bulk motions in the low-density outskirts of the cluster. While these 
bulk motions may provide additional non-thermal ICM pressure support and affect the 
integrated Compton-Y, the gas in these regions have low density and are thus not visible 
in the mock CMB intensity distortion images of Figures~\ref{fig:SZdistortion1} and 
\ref{fig:SZdistortion2} (left panels).

\subsection{SZE Substructure from Shocked Gas}
	
	SZE substructures in the ICM of several clusters uncovered by recent high-resolution 
SZE observations have often been attributed to hot, shocked gas associated with mergers 
\citep[e.g.][]{ko11, plag12}. In our mock SZE images, projection 2 was selected to highlight 
the morphology of the merger shock, as both substructures A and B have strong transverse 
motions in the plane of the images. In the projection 2 view of Figure~\ref{fig:SZdistortion2}, 
substructure A is moving approximately upwards with transverse velocity 1272 km s$^{-1}$, 
and away from the observer with line of sight velocity $-337$ km s$^{-1}$ 
in the rest-frame of the cluster. A large region of disturbed gas surrounds substructure A,
shocked heated by the merger. This is most clearly visible at 90 GHz 
(top left panel of Figure~\ref{fig:SZdistortion2}), and causes the cluster to be asymmetric
in these mock images. 

	In Figure~\ref{fig:SZdistortion2}, substructure B is moving leftwards with transverse velocity 
2493 km s$^{-1}$, and towards the observer with line of sight velocity $94$ km s$^{-1}$ in the 
rest-frame of the cluster. A shock from the merger is visible around substructure B, traveling 
leftwards. These features are also clearly visible in projected temperature map of the system (not shown).
This is reminiscent of the canonical bow shock seen in X-ray observations of the 
Bullet Cluster, in which the merging subcluster has a similar velocity and mass ratio \citep{ma02, sp07}. The 
shock front of substructure B is much sharper than that of substructure A because substructure 
A merged slightly earlier than substructure B. Since substructure A has already traveled into the 
low-density outskirts of the main cluster at this epoch, its shock front in Figure~\ref{fig:SZdistortion2} 
is much less prominent in high-resolution SZE images than substructure B's.

\subsection{Disentangling Bound Subhaloes from the ICM}	
	
	Although we have qualitatively described the effects of the kSZ and merger shocks in our mock 
SZE images, SZE substructures are combinations of a variety of effects and processes. For example,
it is unclear whether the strong kSZ along the line of sight of merging substructures discussed in Section 
4.1 is due to cold, dense gas in their remnant cores or due to fast moving shocked gas. It would be 
particularly helpful for interpreting observed SZE substructures to separate the gravitationally-bound
subhaloes in our simulated cluster from the background ICM. This will allow us to probe their SZE 
signals separately and identify the exact origin of the observable SZE substructures. 
To this end, we run AMIGA Halo Finder \citep[AHF;][]{gi04, kn09} on the main cluster at simulation
time step $z = 0.05$. AHF recursively identifies density peaks in the cluster in a hierarchical tree 
at different resolutions to find haloes and their subhaloes (and sub-subhaloes inside subhaloes, etc.), 
an algorithm well suited for high-resolution cosmological cluster simulations. Once all haloes are isolated, 
AHF assigns each individual particle to a halo by starting at the smallest haloes and 
iteratively removing particles in the halo that are gravitationally unbound, assigning them
instead to their larger parent halo. The resulting haloes thus consists only of bound particles, 
robustly separating subhaloes from parent haloes for analysis. This careful separation of bound and 
unbound particles in each halo is crucial in the dense environments of galaxy clusters, 
where a significant fraction of gas in the cluster galaxies may not be bound to the galaxy, 
but are instead part of the background ICM of the main cluster.

\begin{figure*} 
\centering
\includegraphics[width=0.49\textwidth]{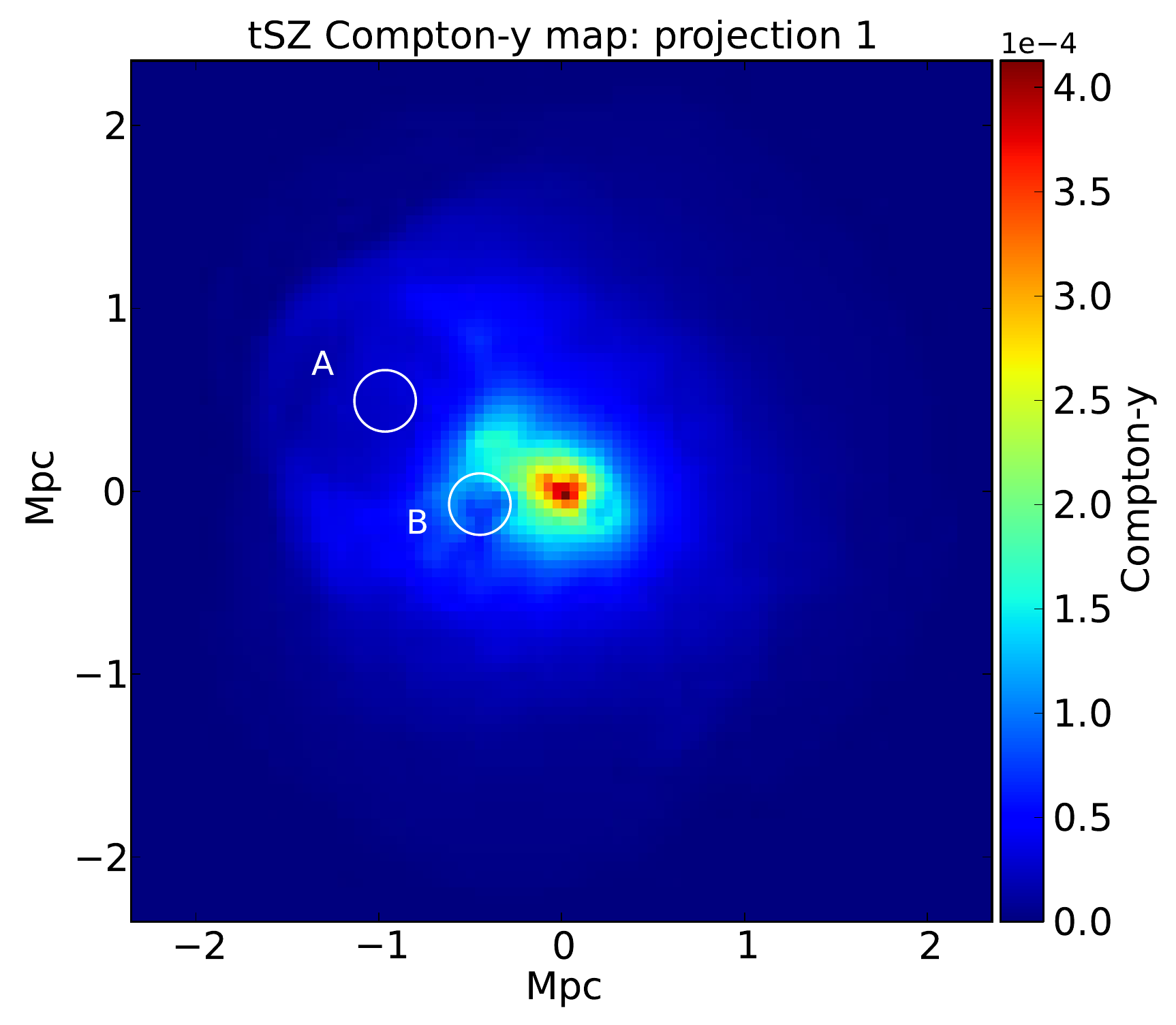} 
\includegraphics[width=0.49\textwidth]{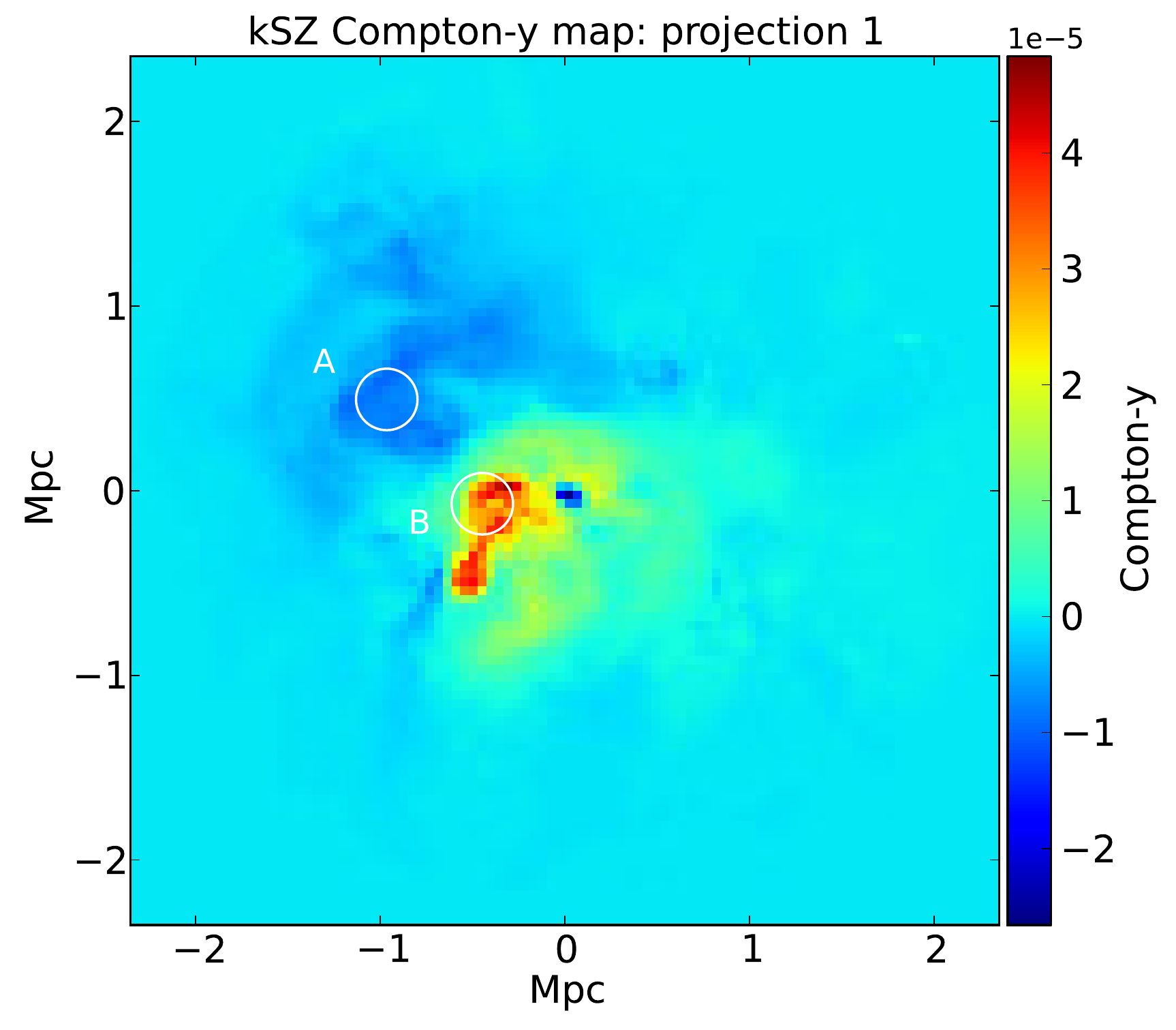}\\ 
\includegraphics[width=0.49\textwidth]{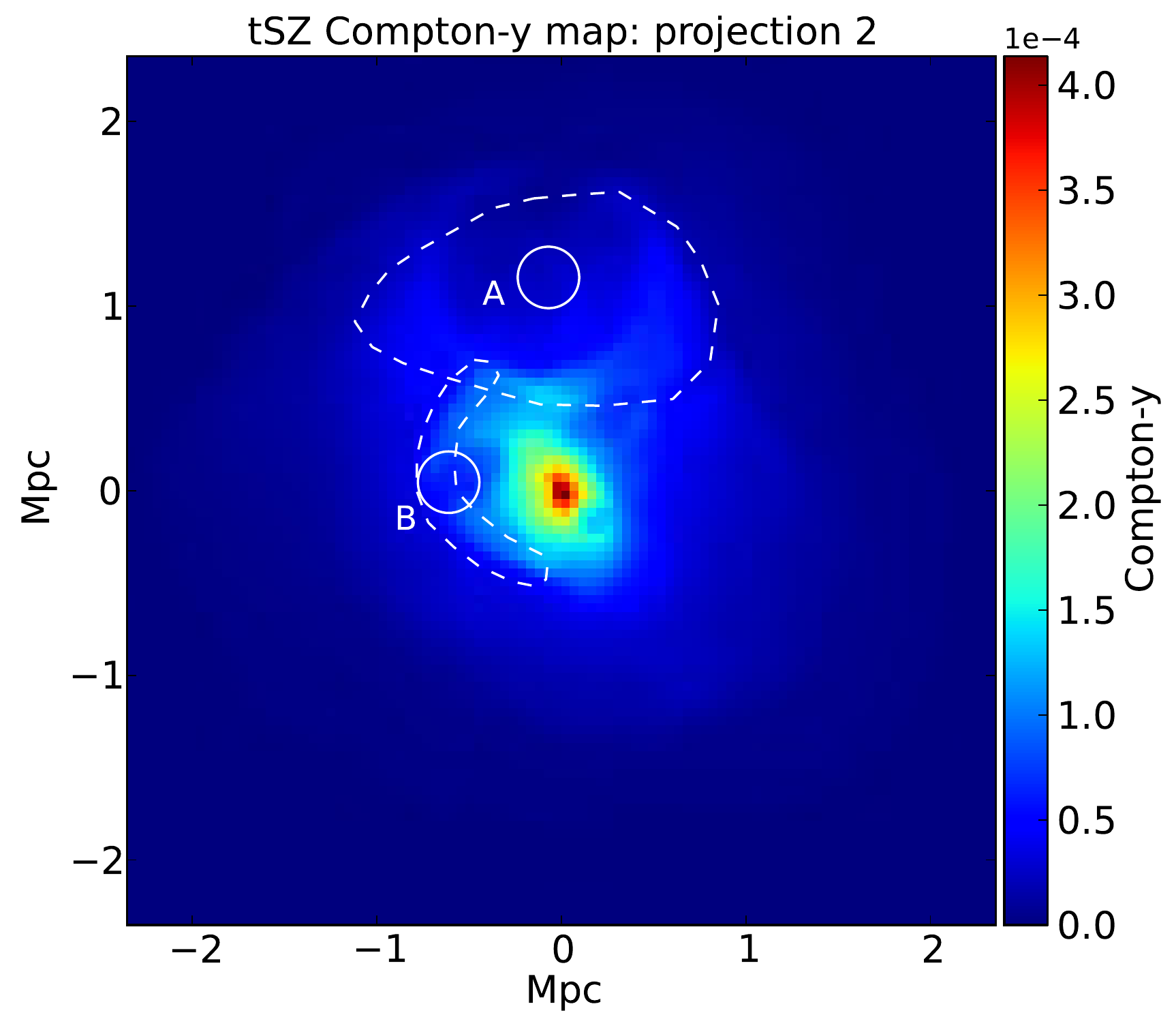} 
\includegraphics[width=0.49\textwidth]{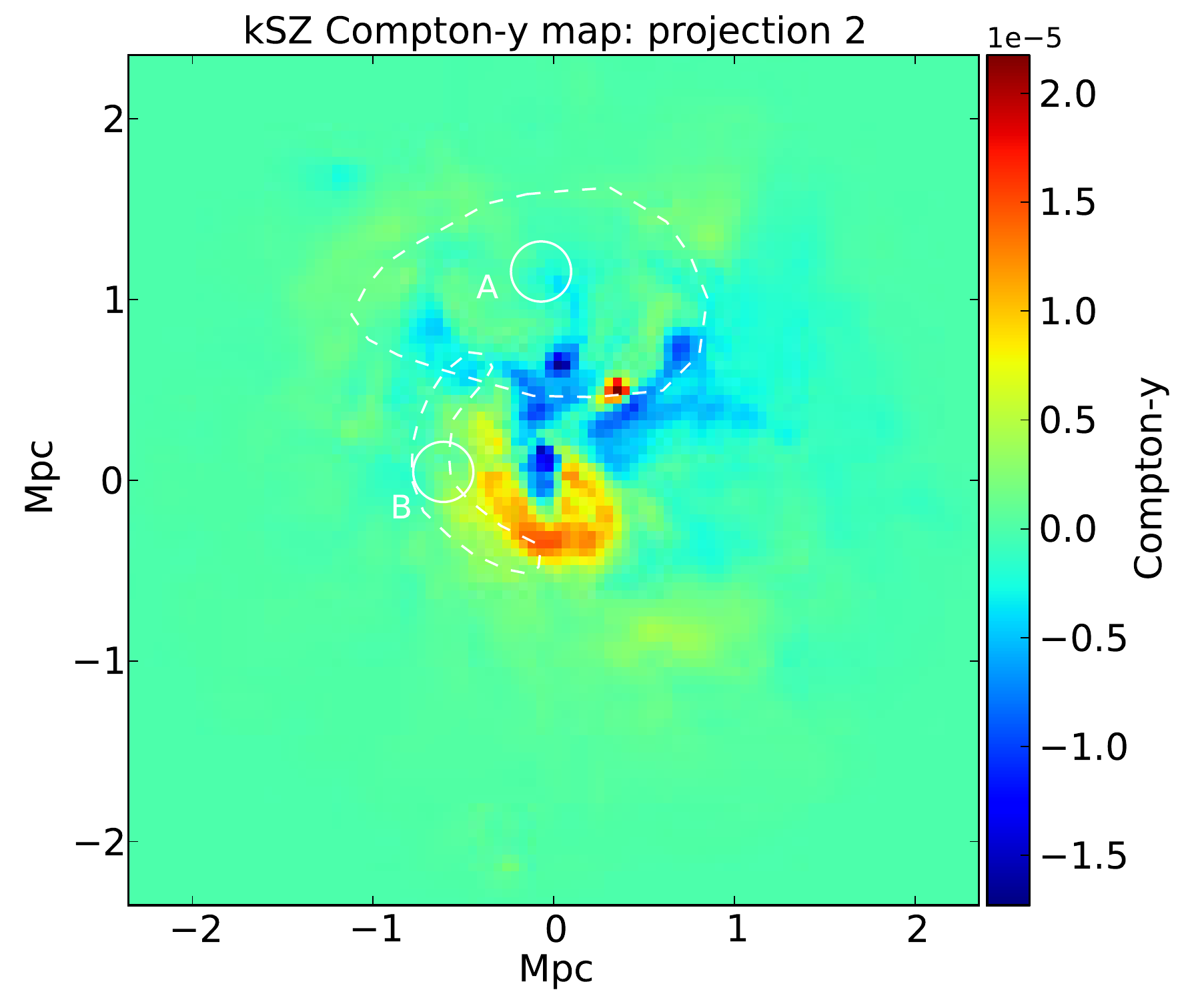} 
\caption{
Compton-y map of the background ICM, gravitationally bound only to the main cluster 
and not to any subhaloes, from the tSZ (left column) and kSZ (right column), along 
projections 1 (top row) and 2 (bottom row). Compared to Figure~\ref{fig:SZcomptony}, 
the kSZ Compton-y around the substructures along projection 1 (top right) are far weaker, 
while the shocked gas around the substructures in the tSZ Compton-y map along 
projection 2 (bottom left) are still faintly visible (encircled with dashed lines). 
This suggests that the observable kSZ is dominated by the gas in the remnant core of the 
substructures rather than the shock, while shocked gas can produce extended 
tSZ structures in the background ICM. 
}
\label{fig:SZcomptony_background}
\end{figure*}

\begin{figure*}
\centering
\includegraphics[width=0.98\textwidth]{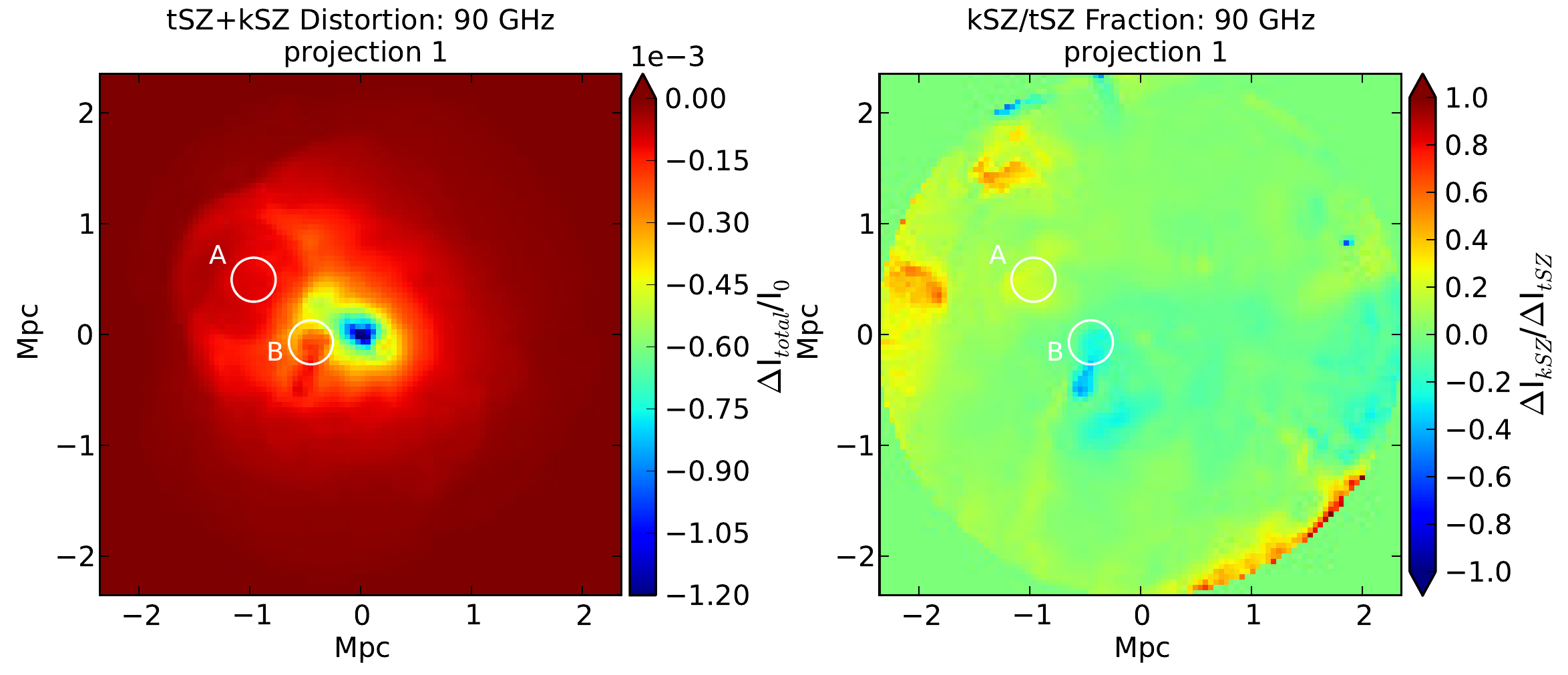}\\
\includegraphics[width=0.98\textwidth]{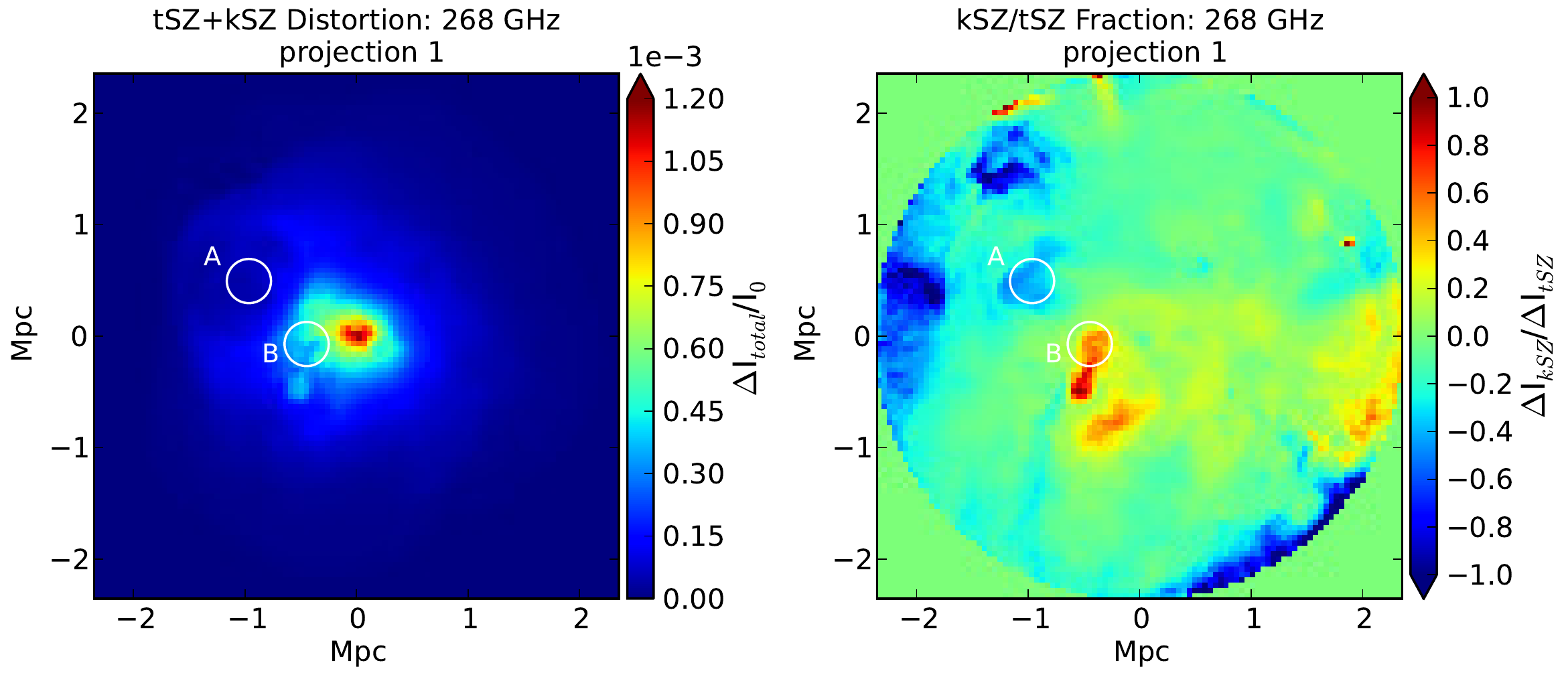}
\caption{
Mock images of the total tSZ+kSZ intensity distortion at 90 GHz and 268 GHz, 
for projection 1 (left column) of the ICM gravitationally bound only to the 
main cluster and not to any subhaloes (compare to Figure~\ref{fig:SZdistortion1}). 
The kSZ/tSZ intensity distortion ratio in each of these maps are also shown (right column).
}
\label{fig:SZdistortion_background1}
\end{figure*}

\begin{figure*}
\centering
\includegraphics[width=0.98\textwidth]{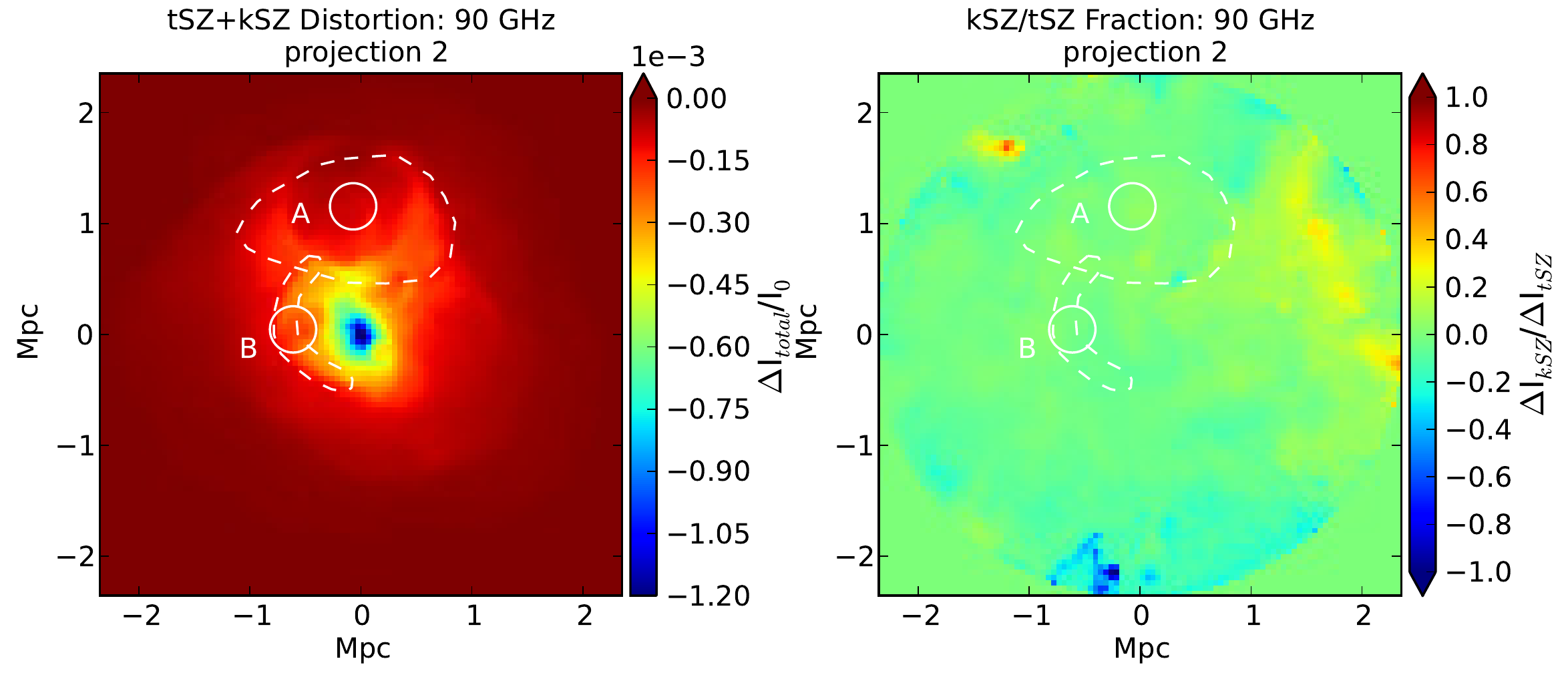} \\
\includegraphics[width=0.98\textwidth]{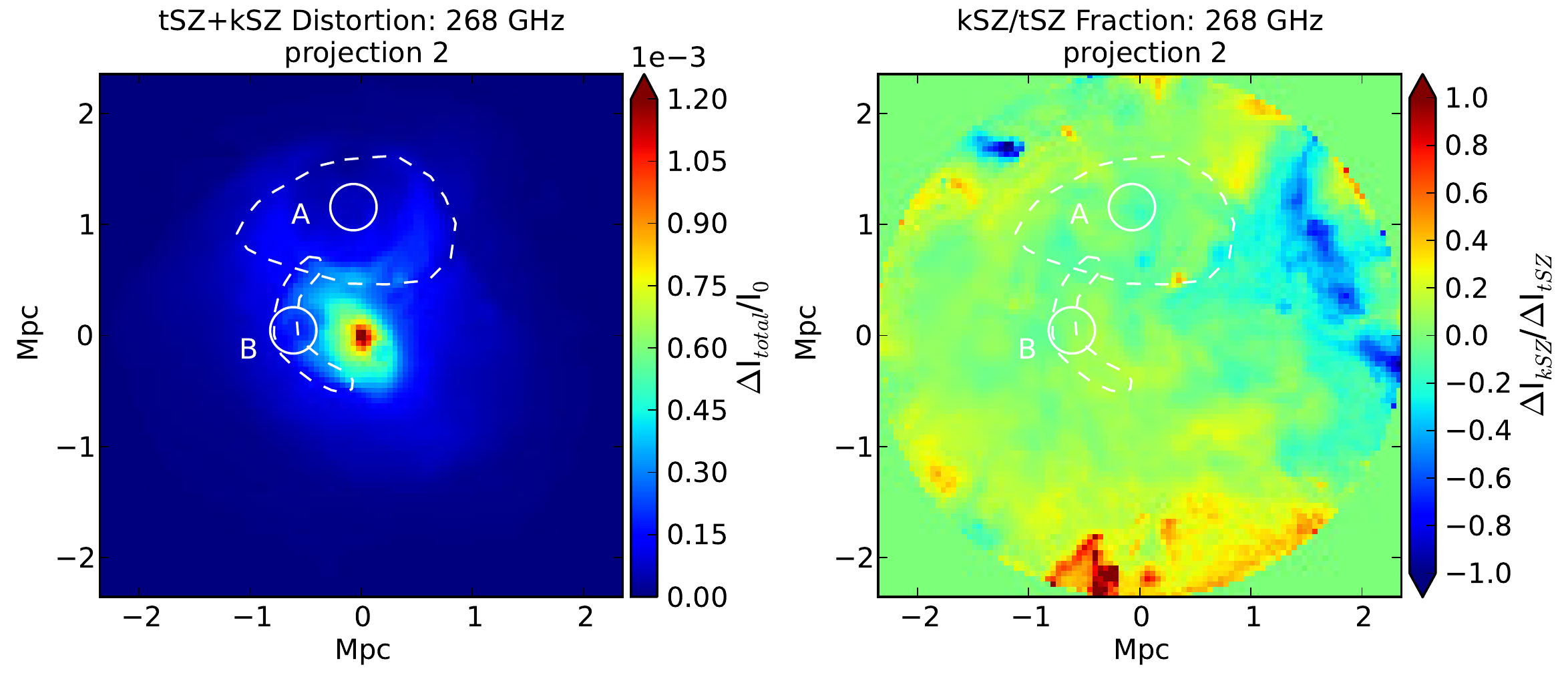}
\caption{
Mock images of the total tSZ+kSZ intensity distortion for the ICM gravitationally bound only 
to the main cluster, similar to Figure~\ref{fig:SZdistortion_background1} but for projection 2
(compare to Figure~\ref{fig:SZdistortion2}). 
}
\label{fig:SZdistortion_background2}
\end{figure*}

	From the haloes identified by AHF, substructures A and B are by far the largest subhaloes 
within the main cluster by more than an order of magnitude, verifying the ability of high-resolution 
SZE observations to detect large subclusters in clusters. In Figure~\ref{fig:SZcomptony_background},
we produce tSZ and kSZ Compton-y maps of just the background ICM, which is gravitationally bound 
only to the main cluster and not to any substructures. In other words, we have reproduced 
Figure~\ref{fig:SZcomptony} after removing all bound substructures. Comparison between 
Figures~\ref{fig:SZcomptony_background} and \ref{fig:SZcomptony} shows that the large high-pressure 
regions corresponding to substructures A and B are no longer present in the background ICM tSZ 
Compton-y maps, and instead there are faint cavities since the volume of space occupied by 
subcluster gas are now empty. The integrated Compton-Y due to the gravitationally-bound gas in 
substructure A accounts for 10\% of the total Y$_{200}$ of the cluster at this epoch, while substructure 
B accounts for only 1\%. Large-scale disturbances in the background ICM due to shocked gas 
are still present in the tSZ Compton-y maps, while some line of sight motions in the background ICM 
around the merging substructures in projection 1 is visible in the kSZ Compton-y maps (especially 
for substructure B). Mock images of the observable CMB intensity 
distortions of this background ICM at 90 GHz and 268 GHz are shown in Figures~\ref{fig:SZdistortion_background1} and \ref{fig:SZdistortion_background2} for projections 1 and 2, 
respectively. Comparison of Figure~\ref{fig:SZdistortion_background1} to Figure~\ref{fig:SZdistortion1}
shows that the strong kSZ effects discussed in Section 4.1 originate in the gravitationally-bound 
gas in the remnant cores, although there is some residual motion in the background ICM gas induced by 
the shock (in addition to other turbulent bulk motions), which we discuss in Section 4.4. 
Comparison of Figure~\ref{fig:SZdistortion_background2} 
to Figure~\ref{fig:SZdistortion2} shows that the bow shock around substructure B and extended 
regions of shocked gas around substructure A are not gravitationally bound to the subclusters, 
and have fainter SZE signals than the remnant cores. 

\subsection{Biases in kSZ Velocity Estimates}

	Velocity estimates of merging subclusters in high-resolution SZE observations have 
usually been estimated by first obtaining X-ray temperature and density maps of the main 
cluster. These X-ray derived values have inherent biases when compared to SZE observations, 
including those that stem from the non-trivial differences in the use of X-ray emission-weighted 
temperatures versus electron density-weighted temperatures \citep{di05}, as well as effects of 
gas clumping. Furthermore, these observations leave temperature fluctuations along the line of 
sight in each pixel poorly constrained, thus essentially assuming a one-temperature approximation 
along any particular line of sight. However, assuming that these values are correct, a X-ray 
derived pseudo-tSZ Compton-y map of the cluster can be calculated and compared to SZE 
Compton-y maps at different frequencies, assuming all spectral distortions are due to the 
tSZ \citep[e.g. as done in][]{mr12}. Disagreements between the SZE Compton-y maps and 
the X-ray pseudo-Compton-y maps will thus likely be due to the kSZ, especially when strongly 
divergent features on these maps follow the general frequency-dependency of the kSZ 
(e.g. if the pseudo-Compton-y under-predicts the observed Compton-y on one side of the 
tSZ null, and over-predicts on the other). In this way, the kSZ substructures can be identified, 
and the spectral distortion region surrounding these kSZ substructures can be fitted with a 
tSZ+kSZ model to determine the line of sight velocity of the merging subcluster.

	Similar to the one-temperature approximation, estimation of the subcluster line of 
sight velocity using the above method also assumes that the gas along the line of sight of 
the merging subcluster has one velocity. Of course, not all the gas along that line of sight 
has the same velocity as the subcluster. The background ICM is well-known to have 
large-scale turbulent bulk motions, but these gas motions will have have random directions, 
and usually smaller velocities. Contribution of the motions of the background ICM to
the kSZ along the line of sight of merging subclusters may thus bias kSZ-based subcluster
velocity estimates in high-resolution observations in ways that are currently unclear. To
probe this bias using a simulation requires comparing the mass-weighted
velocity along the line of merging subclusters to their true velocity as determined by isolating
the subclusters using a halo-finder. 

	Using the gas particles in substructures A and B identified and isolated by AHF, we 
calculate their true particle mass-weighted velocities in the pixels along the centers of their 
remnant cores in projection 1 of Figure~\ref{fig:SZcomptony}. We then compare this true 
velocity of the subclusters to the particle mass-weighted velocity inferred from the kSZ 
distortion in Figure~\ref{fig:SZspectrum}, which essentially adds the kSZ contribution from 
motions in the background ICM gas along the pixel in the calculation. We find that for 
substructure A, the line of sight velocity estimate including the background ICM is only 4\% 
smaller than the true line of sight velocity, while for substructure B it is 12\%. Thus, 
the bias in the estimated velocity due to the background ICM is on the order of $\sim$10\% 
in our simulated cluster. The underestimate of the velocity is larger for substructure B due 
to its closer line of sight position to the core of the main cluster in projection 1, and so the 
denser ambient ICM along its line of sight affects its velocity estimates more significantly. 

	The small magnitude of biases in the velocity estimates for both subclusters is primarily 
due to the large density of the subcluster remnant cores in comparison to the much lower 
density of the background ICM (see discussion in Section 4.3). This causes the kSZ along 
the line of sight of the subclusters to be dominated by the motion of the dense 
gravitationally-bound gas in the merging subcluster, and so the resulting velocity estimation 
is fairly accurate. However, an important secondary factor is the additional contributions to 
the kSZ along the line of sight of the subclusters from bulk gas motion in the background ICM 
induced by the merger. The gas in the background ICM with bulk motion includes
merger-shocked gas, as well as gas stripped away from the merging subcluster
due to ram-pressure stripping. Although the kSZ signal from this motion in the background 
ICM is relatively minor (e.g. the kSZ Compton-y is relatively small around the merging 
subclusters in the top right panel of Figure~\ref{fig:SZcomptony_background} in comparison 
to the same region in Figure~\ref{fig:SZcomptony}), the background ICM gas along the line of 
sight will have velocity vectors in the same direction as the merging subclusters. Thus, the 
background ICM gas will contribute some non-zero kSZ signal that is preferentially in the same 
direction as the merging subcluster, reducing the bias on the inferred velocity of the subcluster. 

	To gauge the effect of the bulk gas motions on our kSZ-inferred substructure velocities, we 
recalculate the kSZ-inferred velocities assuming that the background gas not in the merging 
substructures (but along the line of sight of their remnant cores) have zero velocity, effectively 
removing any kSZ signal from the background ICM. Although this may seem unrealistic since
the background ICM is known to have turbulent bulk motions, these motions not induced by
the merger will have random velocity vectors, and do not substantially contribute to the kSZ 
Compton-y. In this extreme limit of a motionless background ICM, the line of sight velocity estimates 
for substructures A and B are underestimated by 13\% and 25\%, respectively. Thus, the motion 
of the background ICM gas induced by the merger reduces the bias on the velocity estimate 
by a factor of $\sim$2 to 3. This is a promising result for high-resolution SZE observations, 
and shows that the bias in the one-velocity approximation along the line of sight of 
merging subcluster due to the background ICM is not dominant in comparison to other 
sources of bias in observations.

\section{summary and conclusions}

	Using a high-resolution zoomed-in cosmological simulation of a galaxy cluster,
we model the tSZ and kSZ CMB spectral distortion to investigate the observable substructures 
during a 10:3:1 ratio triple merger. We create mock SZE images of the merging cluster 
at the peak of its integrated Compton-Y, where substructures are likely to be most prominent 
and the cluster is most detectable in blind SZE surveys. Our mock images are produced in
two perpendicular projections, the first to highlight the effects of the kSZ, and the second to 
highlight the effects of merger shocks on the background ICM. The two merging subclusters are 
clearly observable in our mock SZE images, with opposing velocities in the main cluster rest-frame. 
Using the AMIGA Halo Finder, we separate the two gravitationally-bound merging substructures 
from the main cluster to investigate the relative contributions of the substructure and main 
cluster ICM gas to the observable SZE in high-resolution observations. In particular, we find the 
following.

\begin{enumerate}
\renewcommand{\theenumi}{(\arabic{enumi})}
\item SZE spectral distortions along the line of sight of merging subclusters can be dominated 
by the kSZ if the subclusters' line of sight velocities are large. Single-frequency observations 
leaving the kSZ unconstrained may lead to misinterpretations.  Multi-frequency observations 
that take advantage of the unique frequency-dependence of the tSZ and kSZ are key to studying 
merging systems. 

\item The remnant cores of merging subclusters are visible in resolved SZE images,
and in the triple merger of our simulated cluster, they contribute 10\% and 1\% to the integrated
Compton-Y.

\item Merger shock features including shock fronts, extended regions of shock-heated gas, and 
strong asymmetry of the cluster due to large disturbed regions can be visible in high-resolution 
SZE images. Future follow-up of SZE-selected clusters will be helpful to identify merging and
disturbed clusters, important for reducing scatter in cluster scaling relations induced by these
systems.

\item kSZ velocity estimates of merging subclusters using a one-velocity approximation for
all gas along the line of sight of the subclusters (including the background ICM)
is generally accurate to $\sim$10\% in our simulated cluster. This accuracy is primarily
due to the dominance of the kSZ along the line of sight by the dense remnant core of the 
merging subcluster, but an important secondary factor is the kSZ contribution from 
bulk motions in the background ICM induced by the merger. Since the merger shock and 
ram-pressure stripped gas will cause bulk motions in the background ICM along the line 
of sight in the same direction as the merging subclusters, it will reduce the velocity bias 
in the one-velocity approximation by a factor of $\sim$2-3.
\end{enumerate}

	Current high-resolution SZE observations have already begun to uncover substructures 
associated with mergers in clusters, and have detected strong kSZ contributions from motions of 
merging subclusters. Our encouraging results indicate that these types of SZE observations
at multiple frequencies can robustly probe both the thermal and dynamic state of the ICM. 
Next-generation high-resolution SZE instruments such as MUSTANG-2 and ALMA's band 1 
will also be able to quickly survey large numbers of clusters with much higher 
sensitivities on a variety of angular scales and resolutions. Aside from their important impact 
on cluster cosmology, these observations could also be interesting probes of other phenomena 
in the ICM, such as those in the central regions of cool core clusters. The ability to separate the 
tSZ and kSZ distortions may allow investigation of the coupling of the central AGN to the 
surrounding ICM, where AGN heating of the cooling core gas could be thermal and/or kinetic in 
nature. Our results offer only a glimpse of the science possibilities with high-resolution SZE 
observations, and further simulations including more baryon physics to complement these 
observations will be necessary.

\section{Acknowledgments}
	The authors wish to thank Sarah R. Loebman for help with using AMIGA-AHF, and the 
anonymous referee for a very thorough review of this paper. 
JJR and TRQ acknowledge support from the NSF AISRP program for development of the simulation 
analysis software, and NSF PHY-0205413  for ChaNGa development. Resources supporting this work 
were provided by the NASA High-End Computing (HEC) Program through the NASA Advanced 
Supercomputing (NAS) Division at Ames Research Center. The authors also acknowledge the Texas 
Advanced Computing Center (TACC) at The University of Texas at Austin for providing HPC resources 
that have contributed to the research results reported within this paper, as well as computing resources 
provided by WestGrid and Compute/Calcul Canada.

\bibliographystyle{mn2e}
\bibliography{ref}  

\label{lastpage}
\end{document}